\documentclass[prd, twocolumn, nofootinbib, notitlepage, superscriptaddress, preprintnumbers]{revtex4-1}

\usepackage{graphicx}
\usepackage{xcolor}
\usepackage{amsmath,bm}
\usepackage[T1]{fontenc}
\usepackage{rotating}
\usepackage{multirow}
\usepackage{hhline}
\usepackage{float}
\floatstyle{plaintop}
\restylefloat{table}

\usepackage{color}
\definecolor{naviBlue}{RGB}{0,0,150}
\usepackage[colorlinks  = true,
            linkcolor   = naviBlue,
            urlcolor    = naviBlue,
            citecolor   = naviBlue,
            anchorcolor = naviBlue]{hyperref}

\newcommand{\gsim}{\ensuremath{\raisebox{-0.4 em}{ $\overset{>}{\sim}$ }}}

\newcommand{\diff}{\mathrm{d}}

\newcommand{\dnds}{$dN/dS$}

\def \apj   {ApJ}
\def \apjs  {ApJS}
\def \apjl  {ApJL}

\def \jcap  {JCAP}
\def \prd   {Phy. Rev. D}

\def \mnras {MNRAS}

\def \aapr  {A.A.Review}

\begin{document}

\preprint{TTK-19-50}

\title{Testing gamma-ray models of blazars in the extragalactic sky}

\author{Silvia Manconi}
\email{manconi@physik.rwth-aachen.de}
\affiliation{Institute for Theoretical Particle Physics and Cosmology, RWTH Aachen University, Sommerfeldstr.\ 16, 52056 Aachen, Germany}

\author{Michael Korsmeier}
\email{michael.korsmeier@to.infn.it}
\affiliation{Institute for Theoretical Particle Physics and Cosmology, RWTH Aachen University, Sommerfeldstr.\ 16, 52056 Aachen, Germany}
\affiliation{Dipartimento di Fisica, Universit\`a di Torino, Via P. Giuria 1, 10125 Torino, Italy}
\affiliation{Istituto Nazionale di Fisica Nucleare, Sezione di Torino, Via P. Giuria 1, 10125 Torino, Italy}

\author{Fiorenza Donato}
\email{fiorenza.donato@to.infn.it}
\affiliation{Dipartimento di Fisica, Universit\`a di Torino, Via P. Giuria 1, 10125 Torino, Italy}
\affiliation{Istituto Nazionale di Fisica Nucleare, Sezione di Torino, Via P. Giuria 1, 10125 Torino, Italy}

\author{Nicolao Fornengo}
\email{nicolao.fornengo@to.infn.it}
\affiliation{Dipartimento di Fisica, Universit\`a di Torino, Via P. Giuria 1, 10125 Torino, Italy}
\affiliation{Istituto Nazionale di Fisica Nucleare, Sezione di Torino, Via P. Giuria 1, 10125 Torino, Italy}

\author{Marco Regis}
\email{marco.regis@to.infn.it}
\affiliation{Dipartimento di Fisica, Universit\`a di Torino, Via P. Giuria 1, 10125 Torino, Italy}
\affiliation{Istituto Nazionale di Fisica Nucleare, Sezione di Torino, Via P. Giuria 1, 10125 Torino, Italy}

\author{Hannes Zechlin}
\email{hzechlin@gmail.com}
\affiliation{Istituto Nazionale di Fisica Nucleare, Sezione di Torino, Via P. Giuria 1, 10125 Torino, Italy}

\begin{abstract}
The global contribution of unresolved gamma-ray point sources to the extragalactic gamma-ray background has been recently measured down to gamma-ray fluxes lower than those reached with standard source detection techniques, and by employing the statistical properties of the observed gamma-ray counts.
We investigate and exploit the complementarity of the information brought by  the one-point statistics of photon counts (using more than 10 years of \emph{Fermi}-Large Area Telescope (LAT) data) and by the recent measurement of the angular power spectrum of the unresolved gamma-ray background (based on 8 years of \emph{Fermi}-LAT data). 
We determine,  under the assumption that the source-count distribution of the brightest unresolved objects is dominated by blazars, their gamma-ray luminosity function and spectral energy distribution down to fluxes almost two orders of magnitude smaller than the threshold for detecting resolved sources.
The different approaches provide consistent predictions for the gamma-ray luminosity function of blazars, and they show a significant complementarity.
\end{abstract}

\maketitle

\section{Introduction}\label{sec::intro}
Since the start of operations of the Large Area Telescope (LAT) on board the \emph{Fermi} satellite \cite{2009ApJ...697.1071A}, 
the gamma-ray sky has become an extremely powerful tool to test the nature of high-energy emissions at all latitudes. 
 The all-sky gamma-ray emission measured by \emph{Fermi}-LAT is typically described in terms of:
 (i) Galactic and extragalactic resolved point-like (and few extended)  sources \cite{Fermi-LAT:2019yla};
(ii) a Galactic diffuse emission, caused by the interaction of cosmic rays with the interstellar
gas and radiation fields \cite{2012ApJ...750....3A};  
(iii) the unresolved gamma-ray background (UGRB) \cite{2012PhRvD..85h3007A,Ackermann:2014usa,2018PhRvL.121x1101A},  which is what remains of the total measured gamma-ray emission after the subtraction of (i) and (ii) (sources that are too faint to be detected individually are defined as \textit{unresolved}).
The Extragalactic Gamma-ray Background (EGB) instead includes all the sources of gamma rays outside the Galaxy which have been resolved, plus the UGRB.
 
The UGRB is statistically isotropic, with tiny  angular fluctuations that have been detected at small angular scales \cite{2012PhRvD..85h3007A, Fornasa:2016ohl, Ackermann:2018wlo}. 
The  anisotropies in the UGRB can be ascribed to the global contribution of one (or more) unresolved populations of point sources. 
In addition to contributions from individual sources, the UGRB contains the contributions from diffuse gamma-rays coming from the interaction of ultra high energy cosmic rays with the intergalactic medium \cite{Kalashev:2007sn}. 
The UGRB could also hide signals of annihilation or decay of dark matter  particles in our Galactic halo, or in outer galaxies  \cite{Ullio:2002pj,Ando:2005xg,DiMauro:2015tfa}: however, these searches are hampered by significant uncertainties \cite{Calore:2013yia}, among which the ones connected to the contribution from astrophysical unresolved point sources playing a major role. 
A residual contamination from the cosmic-ray background is present in the isotropic emission observed by the LAT, being most important at low ($<1$~GeV) and high ($>100$~GeV) energies \cite{Ackermann:2014usa}.

Several source populations contribute to the EGB. At high latitudes,  \emph{Fermi}-LAT has detected blazars, radio galaxies, star forming galaxies (SFGs) and milli-second pulsars \cite{Acero:2015hja,Fermi-LAT:2019yla}. Blazars, a class of active galactic nuclei (AGN), are the most numerous population of individual extragalactic gamma-ray sources \cite{2009ApJ...702..523I,2011ApJ...743..171A,2011PhRvD..84j3007A,2012ApJ...751..108A,2012ApJ...753...45S,Fermi-LAT:2019yla}. 
Depending on the orientation of the relativistic jet of the active galaxy with respect to the observer, AGNs are divided in blazars and misaligned AGNs (mAGNs) \cite{1995PASP..107..803U}. 
Blazars are in turn sub-divided into two categories, depending on the presence of optical emission  lines, radio luminosity and the morphology of the emission: 
BL Lacs  do not present strong emission or absorption features, and have low radio luminosity, which comes predominantly from the center and  the jets, while flat spectrum radio quasars (FSRQs) have  broad emission lines and high radio luminosities which are concentrated in the edge-elongated radio lobes \cite{1995PASP..107..803U,1996MNRAS.281..425M,2017A&ARv..25....2P}. 
They also have different gamma-ray photon indices, softer for the FSRQs ($\sim 2.4$) and harder ($\sim 2.1$) for the BL Lacs \cite{DiMauro:2013xta}. 
We remind that in this context radio galaxies and mAGN can be considered equivalent \cite{1995PASP..107..803U}.
In addition to AGN emission, the mechanism causing the diffuse emission in the Milky Way, such as the interaction of cosmic rays in the interstellar gas and with interstellar radiation fields, is expected to produce gamma rays in SFGs. 
Only few galaxies of this type have been detected so far in gamma rays, e.g. the  M31 and M33 \cite{2017ApJ...836..208A}. 
They are intrinsically faint but numerous \cite{2014JCAP...09..043T}, and the extrapolation of models suggests that they can possibly contribute to the observed UGRB
in a significant way \cite{2011ApJ...733...66I,2012ApJ...753...45S,DiMauro:2013zfa,DiMauro:2013xta, 2014ApJ...796...14C, 2014JCAP...09..043T,DiMauro:2015tfa,Ajello:2015mfa}.
However, the extrapolation of the gamma-ray source count to the unresolved flux regime is based  on correlations to the source count observed at different wavelengths, and consequently suffers from significant uncertainties when applied to derive the count distribution much beyond the resolved flux threshold \cite{Fornasa:2015qua}.

The contribution to the EGB from these gamma-ray source populations can be  quantified by their differential source count distribution \dnds. 
This is the  source number density per solid angle element\footnote{The solid angle $d\Omega$ is omitted in our notation.}, where $N$ is the number of sources in a given flux interval $(S, S+dS)$, and $S$ is the integral gamma-ray flux of a source in an energy bin. The \dnds\ for each source class, in the resolved regime, can be determined through the cataloged point sources. The number of resolved sources is however limited by the detection
efficiency of the survey,  which needs to be estimated for each catalog \cite{DiMauro:2017ing}.
The dissection of the EGB composition is currently complemented  by statistical
methods, able to dig deeper into the unresolved regime.  In fact, analyses employing the statistical properties of the observed gamma-ray
counts map have recently measured 
the contribution from individual sources and the diffuse EGB components, down to  gamma-ray fluxes lower than those obtained with  standard source-detection methods 
\cite{2009PhRvD..80h3504D,Malyshev:2011zi,
  2015JCAP...09..027F,2016ApJS..225...18Z,2016ApJ...832..117L,Mishra-Sharma:2016gis}. 
In particular, in Refs.~\cite{2016ApJS..225...18Z} and \cite{2016ApJ...826L..31Z}
 it was shown that the 1-point probability distribution function (1pPDF) of counts maps serves as a unique tool for precise measurements of the contribution from unresolved point sources to the gamma-ray sky and the EGB's composition. 
Within the 1pPDF analysis, the contribution from unresolved point sources to the EGB has been characterized by fitting the non-Poissonian contribution of sources to the photon counts per pixel, with the prediction computed from a description of the \dnds\
with a generic multiply broken power law (MBPL).  
As shown in \cite{2016ApJ...826L..31Z}, the 1pPDF analysis,  performed with the generic MBPL approach, has the sensitivity to probe the extrapolation of the \dnds\  blazar models in the unresolved flux regime. 

The 1pPDF method can be generalized to include a more physical parametrization of the  \dnds. 
We perform here, for the first time, a fit of \emph{Fermi}-LAT data at latitudes $|b|>30$~deg  with the 1pPDF method using a specific phenomenological model for describing the gamma-ray emission from the blazar population as the dominant contributor.
In combination with this analysis, we consider the two-point angular power spectrum of the UGRB, recently measured on 8 years of {\it Fermi}-LAT  data \cite{Ackermann:2018wlo}. Also in this case, blazars are expected to dominate the anisotropy signal~\cite{Ando:2006cr}, and it has been shown that the gamma-ray angular power spectrum (APS) has the power to  constrain the modeling of the unresolved blazar component~\cite{Ando:2017alx}.

In summary, in this paper we combine the investigation of the blazar component in the gamma-ray extragalactic emission, showing that the 1pPDF and the two-point APS  offer complementary information in the determination of the parameters of blazar models. We then confront these results with the characterization of blazar features we obtain in the resolved regime by using the most recent  catalogs of Refs.~\cite{Fermi-LAT:2019yla,Fermi-LAT:2019pir}. 

The paper is organized as follows.
Section~\ref{sec::blazar_model} describes the model we adopt for the gamma-ray luminosity function and spectral energy distribution  of blazars.
The computation of the relevant observables is outlined in Section~\ref{sec::tech}. Results on the combined analysis of the APS and 1pPDF are presented in Section~\ref{sec::results}. In Section~\ref{sec::conclusion} we summarize our results and main conclusions.

\section{Model for the blazar populations}
\label{sec::blazar_model}
The main aim of the paper is to constrain the model of the gamma-ray emission of blazars at all redshifts 
by applying the 1pPDF and APS analyses. These two observables can be computed from the gamma-ray luminosity function (GLF) and spectral energy distribution (SED) of blazars.
We consider here the model for the GLF and SED derived in Ref.~\cite{Ajello:2015mfa}. 
The authors of Ref.~\cite{Ajello:2015mfa} do not differentiate between the two blazar classes (BL Lacs and FSRQs) since the adoption of a larger sample 
allows for a better determination of the integrated emission from the whole population in the regime of overlapping luminosities. Specifically, we adopt the following decomposition of the GLF $\Phi(L_{\gamma}, z, \Gamma) = dN/dL_\gamma dV d\Gamma$ (defined as the number of sources per unit of luminosity $L_{\gamma}$,  co-moving volume $V$ at redshift $z$ and photon spectral index $\Gamma$) in terms of its expression at $z=0$ and a redshift-evolution function:
\begin{equation}
\Phi(L_\gamma,z,\Gamma)= \Phi(L_\gamma,0,\Gamma) \times e(L_{\gamma}, z),
\label{eq:GLF}
\end{equation}
where $L_\gamma$ is the rest-frame luminosity in the energy range $0.1-100$ GeV, given by $L_\gamma=\int^{100~{\rm GeV}}_{0.1~{\rm GeV}} dE_r\,\mathcal{L}(E_r)$, with:
\begin{equation}
\mathcal{L}(E_r, z,\Gamma) = \frac{4 \pi d^2_L(z)}{(1+z)} E\,\frac{dN}{dE}\, ,
\label{eq:lum}
\end{equation}
$E$ being the observed energy, related to the rest-frame energy $E_r$ as $E_r=(1+z)\,E$. The co-moving volume element in a flat homogeneous Universe is given by $d^2V/d\Omega dz=c\,\chi^2(z)/H(z)$, where $\chi$ is the co-moving distance (related to the luminosity distance $d_L$ by $\chi=d_L/(1+z)$), and $H$ is the Hubble parameter.

At redshift $z=0$, the parametrization of the GLF model is~\cite{Ajello:2015mfa}: 
\begin{eqnarray}
\Phi(L_\gamma,0,\Gamma) &=& \frac{A}{\ln(10) L_\gamma} \left[ \left( \frac{L_\gamma}{L_0} \right)^{\gamma_1} + 
\left( \frac{L_\gamma}{L_0} \right)^{\gamma_2} \right]^{-1} \label{eq:GLF_0} \\
&& \qquad \times
  \exp \left[ - \frac{(\Gamma-\mu(L_\gamma))^2}{2\sigma^2} \right] \;,\nonumber
\end{eqnarray}
where $A$ is a normalization factor, the indices $\gamma_1$ and $\gamma_2$ govern the evolution of the GLF with the luminosity $L_{\gamma}$ and the Gaussian term takes into account the distribution of the photon indices $\Gamma$ around their mean $\mu(L_{\gamma})$, with a dispersion $\sigma$. 
The mean spectral index is allowed to slightly evolve with the luminosity from a value $\mu^\ast$ as~\cite{Ajello:2015mfa}: 
\begin{equation}
\mu(L_\gamma) = \mu^\ast + \beta  
\left[\log \left(\frac{L_\gamma}{\mathrm{erg~s^{-1}}}\right) - 46\right]\, .
\label{eqn:mu}
\end{equation}
Following the results obtained in Ref.~\cite{Ajello:2015mfa}, we adopt the luminosity-dependent density evolution (LDDE):
\begin{eqnarray}
 e(L_{\gamma}, z) =&&  \left[\left(\frac{1+z}{1+z_c(L_\gamma)}\right)^{-p_1(L_\gamma)} \right. \\
 && \qquad \left. + \left(\frac{1+z}{1+z_c(L_\gamma)}\right)^{-p_2(L_\gamma)}\right]^{-1}\nonumber
\end{eqnarray}
with 
\begin{eqnarray}
z_c(L_\gamma) &=& z_c^\ast \cdot (L_\gamma/10^{48})^\alpha,
\label{eq:zc}\\
p_1(L_\gamma) &=& p_1^\ast +\tau \cdot (\log(L_\gamma) -46),
\label{eq:p1}\\
p_2(L_\gamma) &=& p_2^\ast +\delta \cdot (\log(L_\gamma) -46).
\label{eq:p2}
\end{eqnarray}

Concerning the SED, we model it through a double power law:
\begin{equation}
\frac{dN}{dE} = K \left[\left(\frac{E}{E_b}\right)^{\gamma_a} + \left(\frac{E}{E_b}\right)^{\gamma_b}\right]^{-1}\;,
\end{equation}
where we use the prescription of Ref.~\cite{Ajello:2015mfa} for which  $E_b$ correlates with $\Gamma$ according to $\log(E_b/{\rm GeV}) = 9.25 -4.11 \cdot \Gamma$, thus converting the power-law spectrum into a more meaningful spectral shape for blazars. 
Given a SED,  the flux $S(E_{\rm min},E_{\rm max})$ in a given energy interval is obtained as: 
\begin{equation}
S(E_{\rm min},E_{\rm max}) = \int_{E_{\rm min}}^{E_{\rm max}} \frac{dN}{dE}   {e^{-\tau(E\,,z)}} \; dE,
\label{eq:flux}
\end{equation}
where $\tau(E,z)$ describes\footnote{Note that the function $\tau(E,z)$ differes from the parameter $\tau$ in Eq.~\eqref{eq:p1}.}  the attenuation by the extragalactic background light (EBL) \cite{2010ApJ...712..238F}.
The energy flux $S_E (E_1,E_2)$ in a given energy interval is instead:
\begin{equation}
S_E(E_1,E_2) = \int_{E_{\rm 1}}^{E_{\rm 2}} E\; \frac{dN}{dE} \, {e^{-\tau(E\,,z)}} \; dE.
\label{eq:energyflux}
\end{equation}

The GLF and SED models have a large number of free parameters, which in Ref. \cite{Ajello:2015mfa} have been determined by fitting \emph{Fermi}-LAT catalog data, and follow-up observations of blazars. In our analysis we will adopt as free parameters those which grab the dominant dependencies, \emph{i.e.}  the GLF normalization parameter $A$, the central value $\mu^\ast$ for the photon spectral index $\Gamma$, the power-law index $\gamma_1$ that governs the dependence of the GLF at high luminosity and the central values of the power-law indices $p_1^\ast$ and $p_2^\ast$ which set the redshift dependence of the LDDE. All other parameters have been fixed at the values obtained in Ref.  \cite{Ajello:2015mfa}, for definiteness. We have checked both larger (including e.g. also $z_c^\ast$) and smaller sets of free parameters, obtaining that our method is sensitive dominantly to the stated parameters and we will therefore report the results on this set.

\section{The techniques for dissecting the blazar models }
\label{sec::tech}
As mentioned above, in this paper we analyze the 1pPDF, APS and the most recent gamma-ray catalogs and their combined constraining power. In this section, we describe each of these techniques.

\subsection{The 1pPDF photon-count statistics technique}\label{sec::setup1pPDF}
The 1pPDF method relies on defining a probability generating function - generically  derived from a superposition
of Poisson processes - for the photon count maps.
The mathematical formulation of the 1pPDF method, its implementation,
and its application to \emph{Fermi}-LAT data are
discussed in \cite{Malyshev:2011zi,2016ApJS..225...18Z,2016ApJ...826L..31Z}, to which we refer for any detail (see also \cite{Zechlin:2017uzo}). 
In this method, the expected number of point sources in map pixel
$p$ contributing exactly $m$ photons to the total pixel photon content is given  by the
\dnds, being $S$  the integral photon flux of a source in the energy band $[E_\mathrm{min},E_\mathrm{max}]$ (observed energies)
as defined in Eq. (\ref{eq:flux}):
\begin{equation}\label{eq:xm}
x^{(p)}_m = \Omega_\mathrm{pix} \int_0^\infty \mathrm{d}S \,
\frac{\mathrm{d}N}{\mathrm{d}S} \,\frac{[\mathcal{C}^{(p)}\!(S) ]^m}{m!} 
e^{-\mathcal{C}^{(p)}\!(S)} ,
\end{equation}
where $\Omega_\mathrm{pix}$ is the solid angle of the pixel, and $\mathcal{C}^{(p)}\!(S)$ denotes the average number of photons by a source with flux $S$ which contributes to the pixel $p$.
The isotropic distribution of gamma-ray point sources \dnds\ was generically parameterized with a 
  MBPL in Refs. \cite{2016ApJS..225...18Z,2016ApJ...826L..31Z}, with the overall
  normalization, a number of $N_\mathrm{b}$ break positions and
  therefore $N_\mathrm{b}+1$ power-law components connecting the
  breaks as free parameters. In the current analysis we instead progress beyond this generic description, and assess if the \dnds\ of high latitude, extra-galactic sources required to fit the data can be described by 
  a blazar population, described by the physical model of the previous section.
We concentrate our analysis to photon energies in the interval from 1~GeV to 10~GeV, which is where we have at the same time  large statistics and a good angular resolution of the {\it Fermi}-LAT detector.

The differential number of blazars per integrated flux and solid angle \dnds\  can be computed from the model described in Section~\ref{sec::blazar_model} as:
\begin{equation}
\frac{dN}{dS} = \int_{0.01}^{5.0} dz \int_1^{3.5} d\Gamma \, 
\Phi[ L_\gamma(S_{\rm E},z,\Gamma),z,\Gamma] \, \frac{dV}{dz} \, 
\frac{dL_\gamma}{dS},
\label{eq:dnds}
\end{equation}
where $L_\gamma(S_{\rm E},z,\Gamma)$ is the luminosity of a source endowed with an energy flux $S_E $, located at redshift $z$ and with spectral index $\Gamma$, being $S_{\rm E}$ the flux in a specific energy bin. The integration bounds  for $\Gamma$ in Eq. (\ref{eq:dnds}) are such to properly cover the distribution of observed blazars, while the integration bounds for the redshift $z$ cover the interval in which we expect the vast majority of their emission \cite{Ajello:2015mfa}. 

Within the 1pPDF method applied here, the total gamma-ray emission is described by summing an isotropic distribution of point-like blazars and two diffuse background components, the Galactic foreground emission and an additional isotropic component, which
are described by 1-photon source terms.
The total diffuse contribution $x_\mathrm{diff}^{(p)}$ is then given by:
\begin{equation}\label{eq:xdiff}
  x_\mathrm{diff}^{(p)} = A_\mathrm{gal} x_\mathrm{gal}^{(p)}  + \frac{x_\mathrm{iso}^{(p)}}{F_\mathrm{iso}} F_\mathrm{iso}. 
\end{equation}
For the isotropic component $x_\mathrm{iso}^{(p)}$, we use the integral flux $F_\mathrm{iso}$ as a sampling parameter, in order to have physical units of flux\footnote{{We note that that the ratio ${x_\mathrm{iso}^{(p)}}/{F_\mathrm{iso}}$ does not depend on  $F_\mathrm{iso}$.}}.
The first term accounts for the Galactic foreground emission, described
  with an interstellar emission model (IEM). Further details on the considered IEMs are given below. 
The global normalization of the IEM template $A_\mathrm{gal}$ is taken as a free fit nuisance parameter. The second term describes  all contributions
  indistinguishable from purely diffuse isotropic emission. The
  diffuse isotropic background emission is assumed to follow a power
  law spectrum (photon index $\Gamma=2.3$ , see Refs.~\cite{Ackermann:2014usa,2016ApJS..225...18Z}), with its integral flux
  $F_\mathrm{iso}$ serving as the free normalization parameter.
  
 Concerning the data-set, we analyzed all-sky \emph{Fermi}-LAT gamma-ray data  from 2008 August 4
(239,557,417\,s MET) through 2018 December 10 (566,097,546\,s MET). 
We used \texttt{Pass 8} data$^\text{\ref{foot::pass8}}$
\footnotetext{\label{foot::pass8}Publicly available at
  \url{https://heasarc.gsfc.nasa.gov/\\FTP/fermi/data/lat/weekly/photon/}. 
  More details are found at 
  \url{https://fermi.gsfc.nasa.gov/ssc/data/analysis/documentation/Cicerone/Cicerone_Data/LAT_DP.html}
},
along with the corresponding instrument response functions. 
The Fermi Science Tools (v10r0p5)\footnote{\url{https://fermi.gsfc.nasa.gov/ssc/data/analysis/software/}}
were employed for event selection and data processing.
The data selection referred to standard quality
selection criteria (\texttt{DATA\_QUAL==1} and
\texttt{LAT\_CONFIG==1}), to values of the rocking angle of the satellite smaller than $52^\circ$, and maximum zenith angle of $90^\circ$. 
We selected events passing the
\texttt{ULTRACLEANVETO} event class, and we use the corresponding instrument response functions. 
A correction for the source-smearing effects coming from the finite detector point-spread function (PSF) has been also applied in Eq.~\eqref{eq:xm}, as detailed in Ref.~\cite{2016ApJS..225...18Z}.
To avoid significant PSF smoothing, the event sample is restricted to the \texttt{PSF3} quartile (see \cite{2016ApJS..225...18Z,2016ApJ...826L..31Z}). 
Data are analyzed in the energy range from 1~GeV to 10~GeV, and binned using the \texttt{HEALPix} equal-area
pixelization scheme \cite{2005ApJ...622..759G} with a resolution parameter $\kappa=7$, being $N_{\rm pix}=12 N^2_{\rm side}$ the number of pixels, with $N_{\rm side}=2^{\kappa}$.
To avoid significant contamination from the diffuse emission of our Galaxy, we analyzed the data for $|b|>30$ deg.  
The 1pPDF likelihood function is defined as the L2 method in \cite{2016ApJS..225...18Z}. 
The nested sampling algorithm included in the 
\texttt{MultiNest} framework \cite{2009MNRAS.398.1601F} is used to sample the parameter space, 
with 1500 live points together with a tolerance criterion of 0.2.
The IEM has been fixed according to  the official spatial and spectral template as
provided by the \emph{Fermi}-LAT Collaboration for the \texttt{Pass 8}
analysis framework ({\tt gll\_iem\_v06.fits}, see Ref.~\cite{Acero:2016qlg}).

\subsection{The angular power spectrum technique }
\label{sec::Cp}
The APS of the gamma-ray intensity fluctuations is defined as: $C_{\ell}^{ij} = \frac{1}{2\ell+1} \sum_m a_{\ell m}^{i} a_{\ell m}^{j*}$, where the indices $i$ and $j$ label here the energy bins. The coefficients $a_{\ell m}$ are the amplitudes of the expansion into spherical harmonics of the intensity fluctuations, $\delta I_\gamma^i(\vec n) =  \sum_{\ell m} a_{\ell m}^i Y_{\ell m}(\vec n)$, with $\delta I_\gamma^i(\vec n)\equiv I_\gamma^i (\vec n) - \langle I_\gamma^i \rangle$ and $\vec n$ identifies the direction in the sky. The sum defines an average over the modes $m$ for each multipole $\ell$. For $i=j$ the APS describes the energy auto-correlation, while for $i \neq j$ the APS describes the cross-correlation of the fluctuations in two different energy bins.

If the population that dominates the APS is composed of point-like, relatively bright and non-numerous sources, its anisotropy signal is dominated by the so-called Poisson noise term and the APS does not depend on the angular multipole $\ell$, i.e. $C_{\ell}^{ij} \simeq C_{\rm P}^{ij}$. 
One can check that, at the level of fluxes probed by the \emph{Fermi}-LAT, this is the case for blazars \cite{Ando:2006cr}. In our physical model, the blazar APS can be computed as:
\begin{eqnarray}
C_{\rm P}^{ij} &=& \int_{0.01}^{5.0} dz\frac{dV}{dz} \int_1^{3.5} d\Gamma 
\int_{L_{\rm min}}^{L_{\rm max}} dL_\gamma\, \Phi(L_\gamma,z,\Gamma) \label{eq::cp} \\
 &\times& S_i(L_\gamma,z,\Gamma) \, S_j(L_\gamma,z,\Gamma) \left[ 1-\Omega(S_{\rm thr}(L_\gamma,z,\Gamma),\Gamma) \right]\;. \nonumber
\end{eqnarray}
The upper and lower bounds in the $L_{\gamma}$ integration are set to $L_{\rm min}=10^{43}$ erg/s and $L_{\rm max}=10^{52}$ erg/s \cite{Ajello:2015mfa}.
The term $\Omega(S, \Gamma)$ accounts for the \emph{Fermi}-LAT sensitivity to  detect a source, which depends on its photon flux $S$ and spectral index $\Gamma$, and it is described in the next sub-section. 
We will consider both the fourth \emph{Fermi}-LAT catalog (4FGL) of gamma-ray sources~\cite{Fermi-LAT:2019yla} and the third catalog of hard \emph{Fermi}-LAT sources (3FHL)~\cite{TheFermi-LAT:2017pvy}.

The 4FGL catalog is based on eight years of data in the energy range from 50 MeV to 1 TeV and contains 5065 sources which are detected with a confidence level (C.L) above 4$\sigma$. On the other hand the 3FHL catalog is focussed on energies above 10~GeV. It is based on 7 years of data and contains 1556 objects.

The computation of the $C_{\rm P}^{ij}$ requires the same ingredients as in the \dnds\ case: the GLF and SED.
One can interpret the $C_{\rm P}$ as the second moment of the \dnds, as can be seen by comparing Eqs. (\ref{eq:dnds}) and (\ref{eq::cp}). This allows us to combine the constraining power of the 1pPDF method and the anisotropy analysis in the determination of the free parameters characterizing the blazar model. 

The measured $C_{\rm P}$'s adopted in our analysis are taken from Ref.~\cite{Ackermann:2018wlo}, where
the measurement is performed on \texttt{Pass 8} data$^\text{\ref{foot::pass8}}$ of the \texttt{P8R3\_SOURCEVETO\_V2} event class with \texttt{PSF1+2+3} type events.
The data selection comprises 8 years, binned in 12 energy bins between 524~MeV and 1~TeV. 
The contribution from the resolved sources in the energy range $(0.5-14.5)$ GeV, $(14.5-120)$ GeV, $(120-1000)$ GeV is masked using the source list of the FL8Y, FL8Y+3FHL, 3FHL catalogs, respectively \footnote{We note that the $C_P$ measurement of~\cite{Ackermann:2018wlo} is based on the preliminary version of the 4FGL catalog (FL8Y).}. The low latitude Galactic interstellar emission is masked, and a Galactic diffuse template based on the model {\tt gll\_iem\_v6.fits} \cite{Acero:2016qlg} has been subtracted in order to reduce the contamination from high-latitude Galactic contribution. For a full description of methods and results, we refer the reader to Ref.~\cite{Ackermann:2018wlo}. 

The fit of the APS is performed on the auto- and cross-correlation energy bins. The $\chi^2_\mathrm{APS}$ is defined as: 
\begin{eqnarray}
	\label{eqn::chi2_ASP}
	\chi^2_{\mathrm{APS}} = \sum\limits_{i\leq j} \frac{  \left[   \left(   C_P^{ij}    \right)_{\mathrm{meas}} 
	                                                             - \left(   C_P^{ij}    \right)_{\mathrm{th}}     \right]^2   }
	                                                {   \sigma_{C_P^{ij}}^2   } 
\end{eqnarray}
Here the subscript \emph{meas} denotes the measured $C_P$ from Ref.~\cite{Ackermann:2018wlo} and the subscript \emph{th}  denotes the
$C_P$ calculated from Eq.~\eqref{eq::cp}. Furthermore, $\sigma_{C_P^{ij}}^2$ is the uncertainty of the measured $C_P$. 
The likelihood $\mathcal{L} = \exp(-\chi_\mathrm{APS}^2/2)$ is sampled using the \texttt{MultiNest} package in a configuration 
with 2000 live points, an enlargement factor of \texttt{efr=0.7}, and a stopping parameter of \texttt{tol=0.1}. The results
in the next section will be discussed within the frequentist framework. 

\subsubsection{Detection efficiency in the APS analysis }

Let us conclude this section by elaborating more on the issue of the flux threshold sensitivity.
The measurement of the APS is performed by masking sources from the FL8Y and 3FHL catalogs. Therefore, the measured $C_P$ depends on the efficiency of the {\em Fermi}-LAT source detection, see Eq. (\ref{eq::cp}). An exact estimate of such efficiency is challenging, and a typical assumption when calculating the $C_P$ in the blazar model is that this efficiency $\Omega$ abruptly changes from 0 to 1 at a certain flux denoted as $S_{\rm thr}$. 
We adopt such $\Theta$-like cut as our reference model: sources with a given spectral index $\Gamma$ are considered to be undetected ($\Omega=0$)  if their flux in the energy range 1-100 GeV (10-1000 GeV) is below the detection threshold $S_{\rm thr}(\Gamma)$ of the 4FGL (3FHL)\footnote{ We assume that the thresholds of FL8Y and 4FGL are identical.} catalog. 
We define the threshold $S_{\rm thr}$ such that $>98$\% of sources in the catalog with spectral index $\Gamma$ have a flux larger than $S_{\rm thr}(\Gamma)$.
To determine the threshold, the catalog was
binned in $\Gamma$ with bin size equal to 0.1 around $\Gamma=2.3$ and degrading to
0.4 at the extrema of the interval (1 and 3.5), in order to have a sizeable
amount of sources in each bin. We verified that the determination is stable
against changing bin size. We then interpolated the results to build the
function used in the integral of Eq.~\eqref{eq::cp}.
Note that in contrast to many previous analyses we take the $\Gamma$ dependence of $S_{\rm thr}$ into account.

In order to test the impact of our efficiency modeling on the blazar fit, we replace $S_{\rm thr}$ by $k\,S_{\rm thr}$, and marginalize over $k$.
Furthermore, as a test, we replace the $\Theta$-like cut by a smooth function: 
\begin{equation}
 \Omega_\mathrm{smooth} = 1 - \frac{1}{1+(S/S_{\rm thr})^\eta}\;,
\end{equation}
with the parameter $\eta$ varied from 2.5 to 4. 

We have verified that the results of the physical parameters ($A$, $\gamma_1$, $p_1^*$,  $p_2^*$, $\mu^*$) are stable against these changes of the functional form of the efficiency function, with a value for the nuisance parameter $k$ close to 1.

\subsection{The 4FGL and 4LAC catalogs for resolved blazars }
\label{sec::catalog}
As a further technique, relevant for resolved sources, we analyze the most recent source catalogs to constrain the blazar model~\cite{Ajello:2015mfa}. 
The 4FGL catalog \cite{Fermi-LAT:2019yla} is now available, as well as an early release of the fourth catalog of AGNs (4LAC) \cite{Fermi-LAT:2019pir}, both obtained with eight years of data. 
In addition to the blazar type classification, the 4LAC collects also the spectral features, variability and redshift estimates, the last being crucial to constrain the GLF.
The constraints on the GLF obtained from the catalogs of resolved blazars will also be used in Sec~\ref{sec::complementarity} as a prior for the APS fit.

We use the source count distributions extracted from the 4FGL catalog  in a $\chi^2$-fit in which we vary the same five GLF parameters as for the 1pPDF and APS fits: $A$, $\mu^*$, $\gamma_1$, $p_1^*$, and $p_2^*$. 
The total $\chi^2_{\mathrm{4FGL}}$ receives three contributions arising from the total number of observed point sources, the number of associated blazars\footnote{\label{foot:associated}In this paper \emph{associated blazars} refers to the sum of identified and associated sources classified as BL Lacs, FSRQs, or blazars of uncertain type (BCU), 
namely, the 4FGL source classes are \texttt{BLL, BCU, FSRQ, bll, bcu, fsrq}.}, and
blazars with redshift measurements: 

\begin{eqnarray}
	\label{eqn::chi2_4FGL}
	\chi^2_{\mathrm{4FGL}} = \chi^2_{\mathrm{all}} + \max\left( \chi^2_{\mathrm{as}}, \chi^2_{z} \right)
\end{eqnarray}

In the following we will define each contribution. 
For the first term, we extract the source-count distribution of all sources in the 4FGL, 
$(dN/dS)_{\mathrm{all},i}$,
in 12 flux bins $i$ ranging equally spaced in $\log(S)$ from 
$10^{-12}\;\mathrm{cm^{-2}s^{-1}}$ to $10^{-7}\;\mathrm{cm^{-2}s^{-1}}$, where 
 $S=S(1\;\mathrm{GeV},100\;\mathrm{GeV})$.  
To avoid a strong contamination of Galactic sources, 
we restrict the analysis to sources at latitudes with $|b|>30$~deg. 

We compare the extracted source count distribution to the average source count distribution from the 
blazar GLF $\langle dN/dS\rangle_{\mathrm{th},i}$, 
which is the integral of $dN/dS$ (see Eq. (\ref{eq:dnds})) in the energy bin $i$ divided by $\Delta S_i$.
Among the unassociated sources in the 4FGL catalog, we expect that some of them are not blazars. 
Therefore, we use $(dN/dS)_{\mathrm{all},i}$ only as an upper limit in the fit. In terms of
the $\chi^2$ definition this means:
\begin{eqnarray}
	\label{eqn::chi2_4FGL_1}
	\chi^2_{\mathrm{all}} =
	  \sum\limits_i 
	  \begin{cases}
	     \frac{  \left[   \left(       \frac{dN}{dS}      \right)_{\mathrm{all},i} 
	                    - \left\langle \frac{dN}{dS}\right\rangle_{\mathrm{th },i}     \right]^2   }
	          {   \sigma_{\mathrm{all},i}^2   }  & 
	        \!\!\text{if } { \scriptstyle \left\langle \frac{dN}{dS}\right\rangle_{\mathrm{th },i}
	                                > \left(       \frac{dN}{dS}      \right)_{\mathrm{all},i}  }  \\ \, \\
	    0 & \!\!\text{otherwise}
	  \end{cases}~~~
\end{eqnarray}
 The upper limit on the $dN/dS$ adopted in the definition $\chi^2_{\mathrm{all}}$ is complemented with a lower limit arising 
from either the associated sources ($\chi^2_{\mathrm{as}}$) or the sources with redshift measurement ($\chi^2_{z}$). 
It depends on the parameter point
which of the two limits is more constraining. Using the definition of Eq.~\eqref{eqn::chi2_4FGL} we always 
choose the more constraining limit, \emph{i.e.} the one with the larger $\chi^2$.

The contribution of the associated sources is defined with a very similar procedure. There are only two small differences: 
(i) instead of extracting the total source count distribution, we extract the source count distribution of associated 
blazars\href{foot:associated}{$^{\ref*{foot:associated}}$}, $(dN/dS)_{\mathrm{as},i}$, and
(ii) we use $(dN/dS)_{\mathrm{as},i}$ as a lower limit in the fit since the association in the catalog 
might be incomplete. As before, we consider the flux $S=S(1\;\mathrm{GeV},100\;\mathrm{GeV})$. 
The $\chi_\mathrm{as}^2$ is defined by:
\begin{eqnarray}
	\label{eqn::chi2_4FGL_2}
	\chi^2_{\mathrm{as}} = 
	  \sum\limits_i 
	  \begin{cases}
	    \frac{  \left[   \left(       \frac{dN}{dS}      \right)_{\mathrm{as},i} 
	                   - \left\langle \frac{dN}{dS}\right\rangle_{\mathrm{th},i}     \right]^2   }
	         {   \sigma_{\mathrm{as},i}^2   }  & 
	        \!\!\text{if } { \scriptstyle \left\langle \frac{dN}{dS}\right\rangle_{\mathrm{th},i}
	                                < \left(       \frac{dN}{dS}      \right)_{\mathrm{as},i}  } \\ \, \\
	    0 & \!\!\text{otherwise}
	  \end{cases}~~~
\end{eqnarray}

We exploit the redshift information from the 4LAC catalog to constrain the LDDE function by extracting
the source count distribution in 4 redshift bins, $j$:
[0, 0.5], [0.5, 1.2], [1.2, 2.3] and [2.3,4].
The source count distribution, 
$(dN/dS)_{\mathrm{z},ij}$, is extracted equivalently to the procedure described above. The only difference is
that the number count is restricted to the redshift in each bin. 
The corresponding source count distribution of the GLF model, $\langle dN/dS\rangle_{\mathrm{th},ij}$, is obtained
by restricting the integration range of $z$ in Eq.~\eqref{eq:dnds} to the redshift bin.
Since the redshift measurements in the catalog are incomplete, the source count distributions extracted  
from the 4LAC catalog are taken as lower limits: 
\begin{eqnarray}
	\label{eqn::chi2_4FGL_3}
	\chi^2_{z} = 
	  \sum\limits_{i,j} 
	  \begin{cases}
	   \frac{  \left[   \left(       \frac{dN}{dS}      \right)_{   z        ,ij} 
	                  - \left\langle \frac{dN}{dS}\right\rangle_{\mathrm{th},ij}     \right]^2   }
	        {  \sigma_{\mathrm{z},ij}^2   } & 
	        \!\!\text{if } { \scriptstyle    \left\langle \frac{dN}{dS}\right\rangle_{\mathrm{th},ij}
	                                     < \left(       \frac{dN}{dS}      \right)_{   z        ,ij}  } \\ \, \\
	    0 & \!\!\text{otherwise.}
	  \end{cases}~~~
\end{eqnarray}
 Note that by taking as an upper limit on the $dN/dS$ this definition of $\chi^2_{\mathrm{all}}$ and then either $\chi^2_{\mathrm{as}}$ or $\chi^2_{z}$ as a lower limit, there is no double counting in Eq. \eqref{eqn::chi2_4FGL}. 
We have cross checked that the combination of $ \chi^2_{\mathrm{all}} + \chi^2_{z}$ allows us to mostly determine $p_1^*$ and $p_2^*$, 
while the combination of  $\chi^2_{\mathrm{all}} + \chi^2_{\mathrm{as}} $ constrains $A$, $\gamma_1$ and, mildly, $\mu^*$.

In order to sample the 5-dimensional parameter space we use the \texttt{MultiNest} package. We adopt 2000 live points, an enlargement factor of \texttt{efr=0.7}, and a stopping parameter of \texttt{tol=0.1}. 
The results presented in the next section are interpreted in the frequentist approach.

\subsubsection{Detection efficiency in the catalog analysis}
Finally, we discuss the assumptions adopted for the detection efficiency in the analysis of catalog sources.
As described above, the sensitivity to detect point sources in the 4FGL catalog drops below some 
threshold flux, $S_\mathrm{thr}(\Gamma)$. At fainter fluxes the observed source count distribution also drops and its description becomes more cumbersome. 
Since in the catalog analysis we are not splitting sources in bins according to their spectral index, we define a unique $S_\mathrm{thr}$.
We conservatively restrict the sum over $i$ in Eq.~\eqref{eqn::chi2_4FGL_1} to those 
flux bins which are above the maximal threshold flux, determined as described in Section~\ref{sec::Cp}. The latter is $S_\mathrm{thr} = 1.1 \times 10^{-10} \mathrm{cm^{-2} s^{-1}}$ (corresponding to $\Gamma \sim 2.3$).

Note that we do not require to restrict the sums 
in Eqs.~\eqref{eqn::chi2_4FGL_2} and ~\eqref{eqn::chi2_4FGL_3} because they serve as lower limit and a decrease 
of the observed source count distribution only weakens the limit.

\section{Results}
\label{sec::results}
In this section, we first present the results obtained by applying the 1pPDF analysis to the specific blazar \dnds\ model introduced in Section~\ref{sec::blazar_model}. 
Then, we probe the blazar model through the APS analysis, and combine the two methods. 
Finally, we check the compatibility of our results with the 4FGL catalog.

\subsection{Results from the photon-count statistics analysis}\label{sec::results1pPDF}
\begin{figure}[t]
\includegraphics[width=1.0\linewidth]{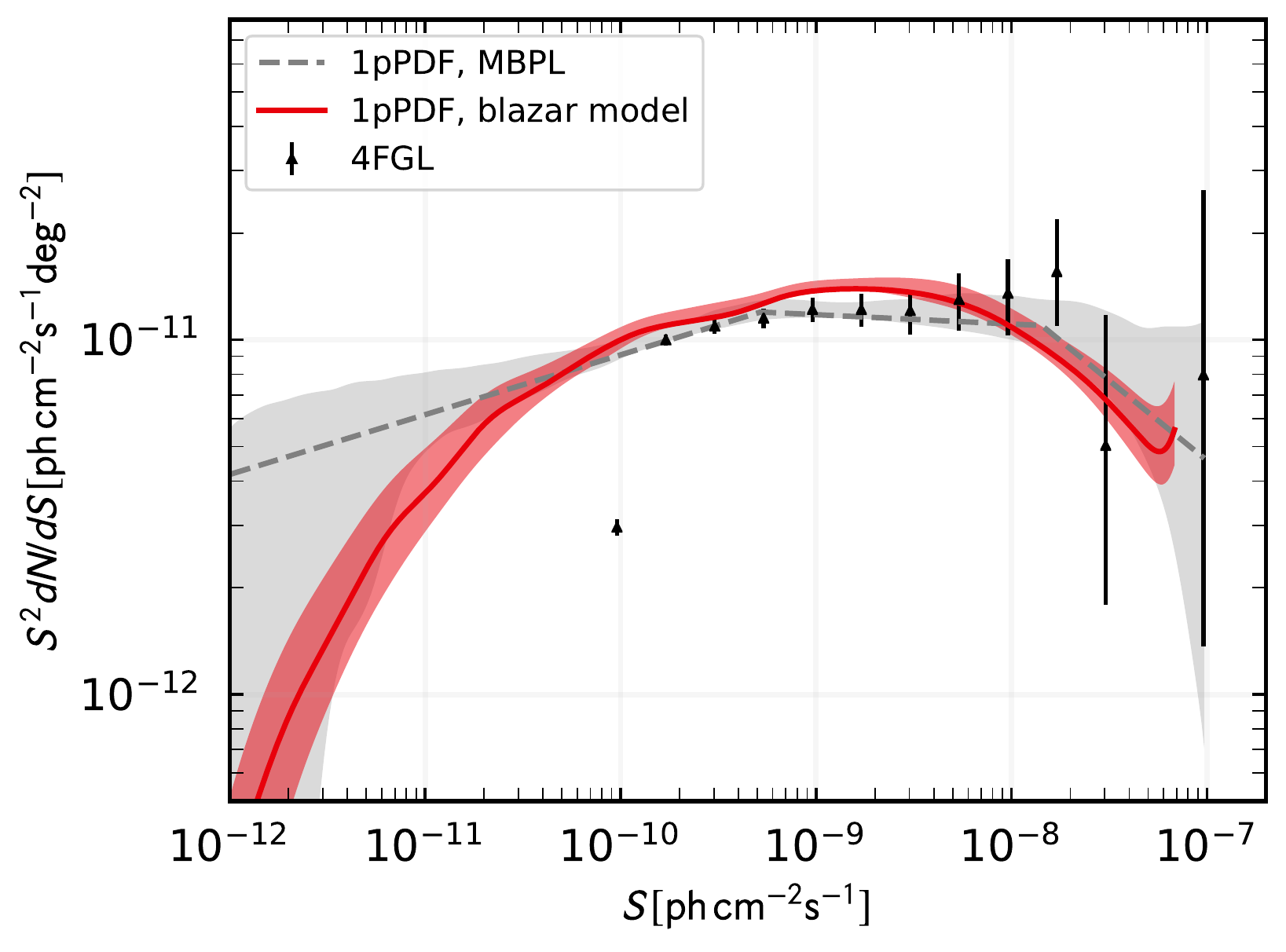}
\caption{Source-count distribution \dnds\ determined by fitting the 
blazar model described in Section~\ref{sec::blazar_model} (red solid line) and using a MBPL parametrization (gray dashed line) with the 1pPDF in the energy range [1-10] GeV. The shaded bands show the 1$\sigma$ uncertainty.	           
The resolved sources from the 4FGL catalog are also displayed.}
\label{fig::fit_results_1ppdf}
\end{figure}

\def \fn_tab {\hyperref[foot::table_best_fit]{$^{\ref*{foot::table_best_fit}}$}}
\begin{table*}[t]
\caption{Best-fit parameters for each of the techniques investigated in this paper. The first column lists the free parameters, while the following four columns contain the corresponding best fits. The last column reports the reference values from Ref.~\cite{Ajello:2015mfa}.}
\centering
\begin{tabular}{ c @{\hspace{10px}} c @{\hspace{10px}} c @{\hspace{10px}} c @{\hspace{10px}} c @{\hspace{20px}} c } \hline\hline
\textbf{Parameter}                                         &   \textbf{1pPDF}                        &  $\bm{C_P}$                        &    \textbf{4FGL}                          &    $\bm{C_P}$\textbf{+4FGL}              &   \textbf{Ref.~\cite{Ajello:2015mfa}}      \\ \hline
$\log_{10}(A/\mathrm{Mpc^{-3}})$                           &   $-8.98^{+0.86}_{-0.49}$               &  $-7.55^{+0.54}_{-5.60}$           &  $-9.10^{+0.37}_{-0.18}$                 &   $-8.89^{+0.08}_{-0.16}$                &   $-8.71^{+0.36}_{-0.47}$                  \\
$\gamma_{1}$                                               &   $0.652^{+0.44}_{-0.02}$               &   $0.36^{+0.17}_{-0.23}$           &    $0.61^{+0.18}_{-0.13}$                 &    $0.56^{+0.07}_{-0.03}$                &   $0.50^{+0.14}_{-0.12}$                   \\
$p_1^*$                                                    &   $3.26^{+2.74}_{-2.26}$                &   $4.89^{+0.11}_{-0.75}$           &    $2.28^{+1.52}_{-1.27}$                 &   $3.32^{+0.99}_{-1.35}$                &   $3.39^{+0.89}_{-0.70}$                   \\
$p_2^*$                                                    &   $-17.5^{+8.60}_{-2.54}$               &  $-19.5^{+7.36}_{-0.50}$           &   $-4.53^{+3.21}_{-1.42}$                 &   $-5.44^{+1.46}_{-0.74}$               &   $-4.96^{+2.25}_{-4.76}$                  \\
$\mu^*$                                                    &   $1.78^{+0.34}_{-0.22}$                &   $2.32^{+0.05}_{-0.09}$           &    $1.93^{+0.89}_{-0.93}$                 &   $2.30^{+0.03}_{-0.04}$                &   $2.22^{+0.03}_{-0.02}$                   \\ \hline
$A_\mathrm{gal}$                                           &   $1.05^{+0.01}_{-0.01}$                &   -                                &    -                                      &    -                                     &   -                                        \\
$F_\mathrm{iso}\;[10^{-7}\mathrm{cm^{-2} s^{-1} sr^{-1}]}$ &   $1.18^{+0.11}_{-0.12}$                &   -                                &    -                                      &    -                                     &   -                                        \\
$k$                                                        &   -                                     &   $0.59^{+0.82}_{-0.09}$           &   -                                      &    $1.29^{+0.13}_{-0.19}$                &   -                                        \\ \hline
-                                                          &   ln$(\mathcal{L})$= -245276.1          &   $\chi^2/\mathrm{dof}$ =80.2/72       &    $\chi^2/\mathrm{dof}$ = 3.2/2 \fn_tab   &   $\chi^2/\mathrm{dof}$ = 90.9/79      &   -                                        \\
\hline\hline
\end{tabular}
\label{tab::bestfit_parameter2}
\end{table*}

The results on the determination of the \dnds\  for high latitude blazars, obtained with the 1pPDF analysis, are shown in Fig.~\ref{fig::fit_results_1ppdf}: the red solid line is the result obtained by using the blazar model of  Section \ref{sec::blazar_model}, while the dashed gray line refers to the results obtained by employing a MBPL, as done in 
\cite{2016ApJS..225...18Z}. The shaded areas of corresponding color denote the $1\sigma$ frequentist uncertainty.  For the physical blazar model of Section \ref{sec::blazar_model} we vary the parameters 
$A$, $\mu^\ast$, $\gamma_1$, $p_1^\ast$ and $p_2^\ast$ and marginalize over two nuisance parameters, the normalization
of the Galactic foreground emission $A_\mathrm{gal}$, and the flux of the isotropic gamma-ray  emission, $F_\mathrm{iso}$. In the case of the MBPL, we adopt a mode with three nodes (see \cite{2016ApJS..225...18Z} for details) and we obtain the following results:  for the normalization parameter $A_\mathrm{S} = 2.31^{+7.67}_{-1.22}  \times 10^9$ cm$^2$ s sr$^{-1}$; 
$S_{b1}= 1.43^{+3.57}_{-0.93} \times 10^{-8}$ cm$^{-2}$ s$^{-1}$, 
$S_{b2}= 5.2^{+8.08}_{-2.94} \times 10^{-10}$ cm$^{-2}$ s$^{-1}$,
$S_{b3}= 2.21^{+97.7}_{-1.18} \times 10^{-13}$ cm$^{-2}$ s$^{-1}$ for the position of the breaks; 
$n_1 = 2.45^{+0.78}_{-0.48}$, $n_2 = 2.03^{+0.10}_{-0.10}$, $n_3 = 1.83^{+0.14}_{-0.15}$, $n_4 = -0.32^{+2.18}_{-1.68}$ for the power-law exponents. The position of the  third break, and the corresponding index $n_4$ at very low fluxes, is not statistically significant.
Finally, Fig.~\ref{fig::fit_results_1ppdf} also shows the counts for all the resolved sources listed in the 4FGL catalog. For each source, the photon flux in the energy bin [1,10]~GeV was calculated by integrating the spectrum obtained by the best-fit spectral model given by the 4FGL catalog, as detailed in Appendix B of Ref.~\cite{2016ApJS..225...18Z}.

The MBPL result shows that the 1pPDF is able to determine the behavior of the \dnds\ more than one order of magnitude in flux lower than the catalog threshold ($S \sim 2-3\times 10^{-10}$ cm$^{-2}$ s$^{-1}$), namely at $S \sim 8 \times 10^{-12}$ cm$^{-2}$ s$^{-1}$,
below which the uncertainty band increases significantly. When this is translated to the physical blazar model, it allows to determine and trust the behavior of the \dnds\  down to 
the same flux level, therefore extending the understanding of the blazar model in the unresolved regime. Let us also notice that the fact that the results obtained with the physical blazar model are very well consistent with those obtained with the generic MBPL analysis and with the 4FGL catalog sources, reinforcing our assumption that point sources
emitting photons at high latitudes in the energy range from 1~GeV to 10~GeV are consistent with a blazar origin even in the unresolved regime.

The best-fit values of the relevant parameters of the GLF blazar model, together with their uncertainties, are reported in Tab. \ref{tab::bestfit_parameter2}.  We obtain values which are largely compatible (except for $p_2^\ast$, where compatibility is present only at about the $2\sigma$ level) with the reference model  of Ref. ~\cite{Ajello:2015mfa}, which was adapted to the resolved component and to a source catalog predating the 4FGL.
In Tab. \ref{tab::bestfit_parameter2} we also show the results for the same parameters, obtained by fitting the 4FGL catalog (see Sec.~\ref{sec::complementarity} and Fig.~\ref{fig::fit_results_4FGL}), in which case the agreement between our results and Ref. \cite{Ajello:2015mfa} is well inside $1\sigma$ for all parameters. These results indicate that the unresolved blazar component (down to fluxes of the order of about $8 \times 10^{-12}$ cm$^{-2}$ s$^{-1}$) has similar properties as those which are currently resolved, with some faint hint of transition relative to the high-redshift  dependence (encoded in $p_2^\ast$).

The photon-count statistics analysis decomposes the total gamma-ray emission at $|b|>30$~deg according to the method outlined in Sec.~\ref{sec::setup1pPDF}. The fractional contributions to the total integral flux $F_{\rm tot}$ \cite{2016ApJS..225...18Z} of each component in the energy bin $[1,10]$~GeV, and for the fit with the blazar model, are found to be: $q_{\rm ps}=0.195^{+0.009}_{-0.005}$ for point sources, $q_{\rm gal}=0.706 \pm 0.004$ for the Galactic diffuse emission, and $q_{\rm iso}=0.084 \pm 0.008$ for the diffuse isotropic background. As for the MBPL fit, we find $q_{\rm ps}=0.247^{+0.018}_{-0.039}$, $q_{\rm gal}=0.705 \pm 0.005$ and $q_{\rm iso}=0.046^{+0.051}_{-0.018}$.

The two nuisance parameters $A_{\rm gal}$ and  $F_\mathrm{iso}$ are statistically well constrained within the 1pPDF fits. We observe a mild degeneracy between the normalization of the point sources (both for the MBPL and the blazar fit) and the diffuse isotropic component $F_\mathrm{iso}$. 
However, as demonstrated by the Monte Carlo validation of the method included in Ref.~\cite{2016ApJS..225...18Z}, the method reconstructs the source-count distribution down to the quoted sensitivity, below which point sources become indistinguishable from a purely isotropic emission.

\subsection{Results from the angular correlation analysis}
\label{sec::results_Cp}
\begin{figure*}[t!]
	\includegraphics[width=0.48\textwidth, trim={1.5cm 0 2.5cm 0}, clip]{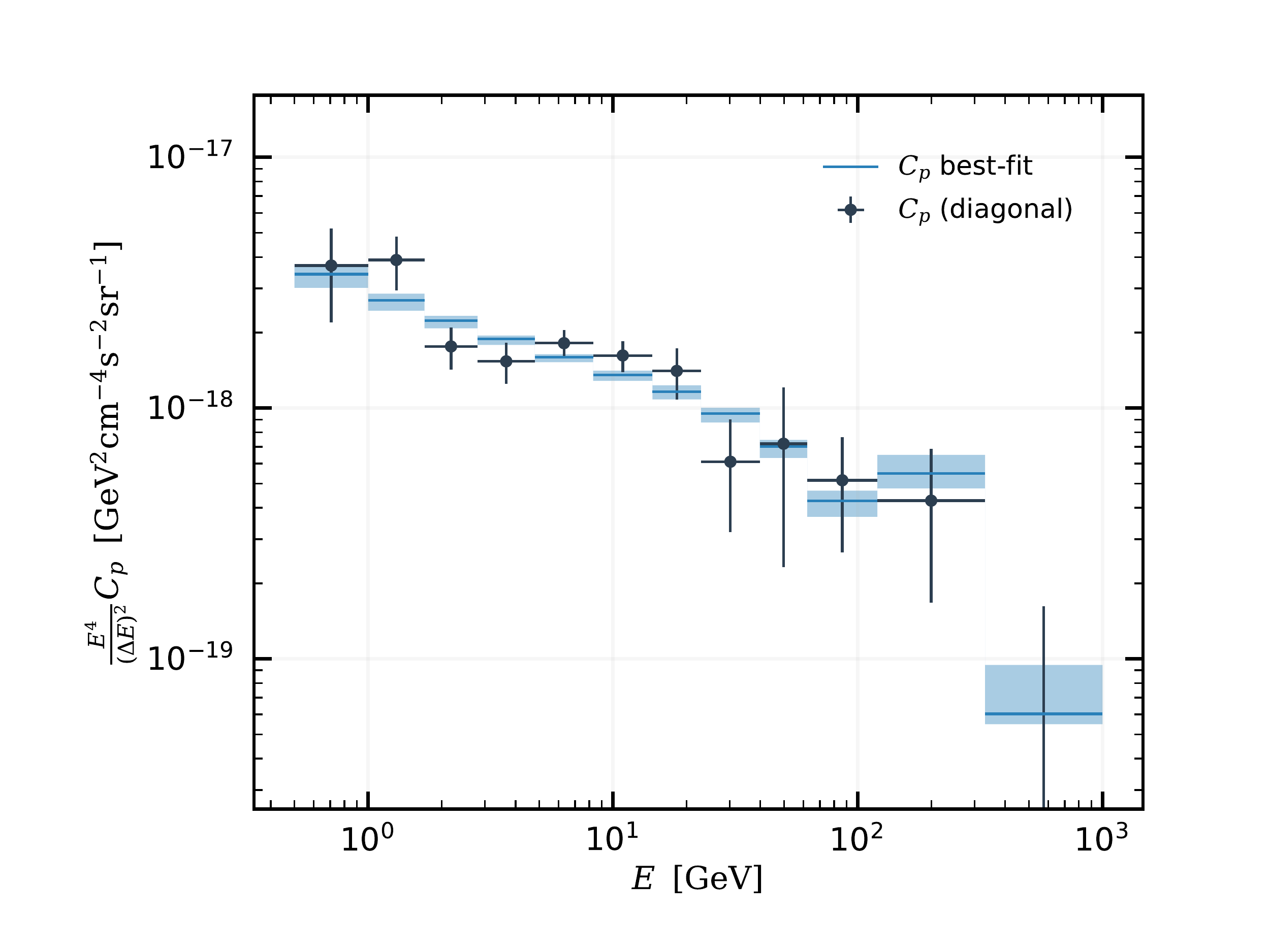}\hspace{0.04\textwidth}\includegraphics[width=0.48\textwidth, trim={1.5cm 0 2.5cm 0}, clip]{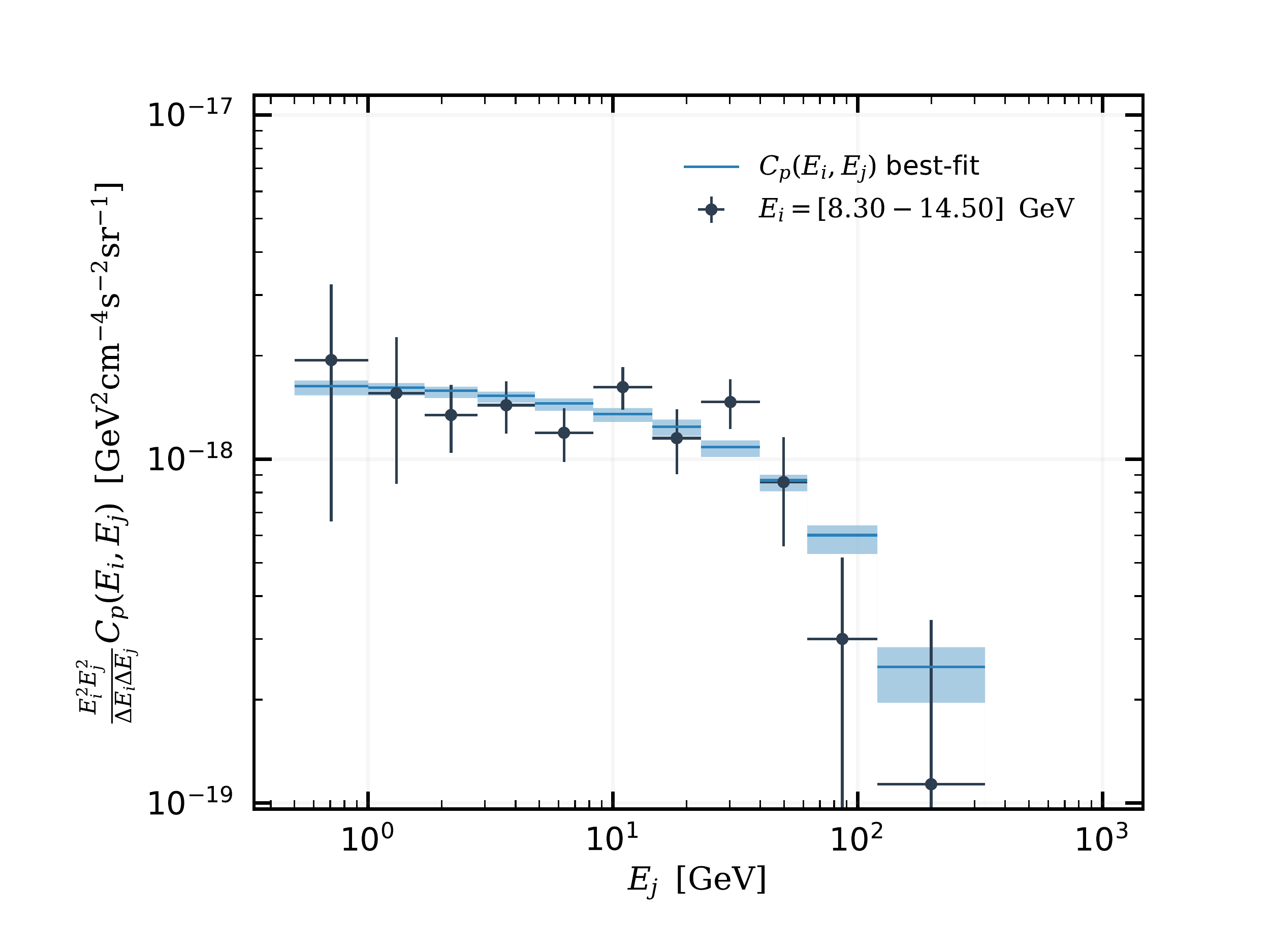}
    \caption{Best-fit result of the blazar model 
	           to the angular correlations amplitude $C_P$ as a function of the energy, as measured in Ref.~\cite{Ackermann:2018wlo}.
	           The left panel refers to the autocorrelation (in energy), while the right panel shows one set of cross-correlations (in energy) of one selected energy bin
	           (8.3~GeV--14.5~GeV) with all others. The shaded bands display the 1$\sigma$ (frequentist) uncertainty.
	}
  \label{fig::fit_results_Cp}
\end{figure*}
%
\begin{figure*}[t!]
	{\includegraphics[width=0.9\textwidth]{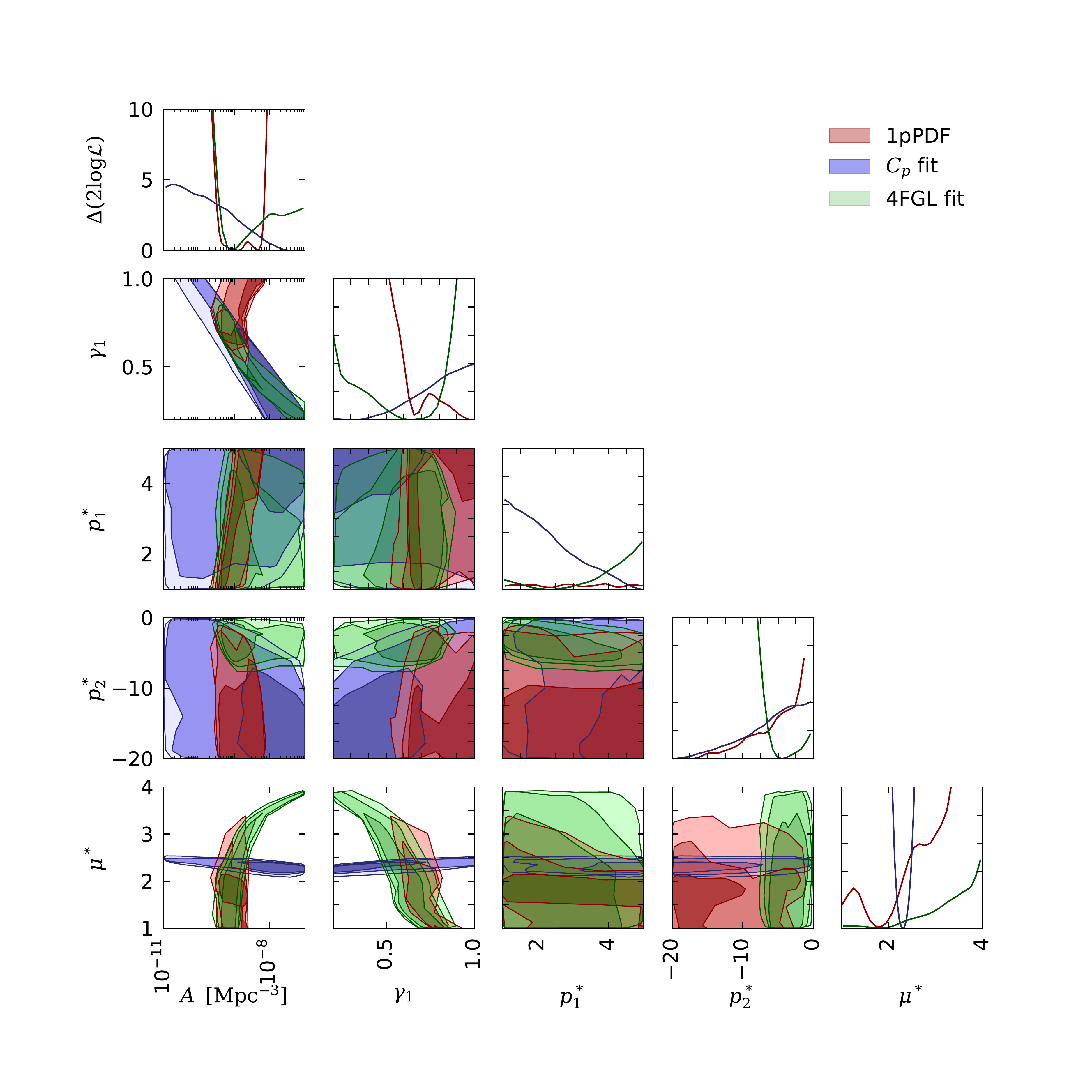}}
        \caption{Constraints on 
 the blazar model parameters obtained by fitting the source count 
                  distribution using the 1pPDF method (red) and the angular correlations amplitude $C_P$
                  (blue). The boxes on the diagonal show the likelihood profile for each of the fit parameters (the
                  vertical axis of each box always spans from 0 to 10 in linear scale), while the other panels 
                  show the 1, 2 and 3$\sigma$ C.L. contours of the 2-dimensional $\chi^2$ distribution
                  for each combination of the parameters.
                }
  \label{fig::triangle_1ppdf_Cp}
\end{figure*}

In the APS fit, we consider the auto- and cross-correlation measurements involving all the energy bins from 0.5~GeV to 1~TeV adopted in Ref.~\cite{Ackermann:2018wlo}.
The number of energy bins is $N_b=12$, and so of auto-correlation data, while the number of cross-correlation measurements is $N_b\times(N_b-1)/2=66$.

In this analysis,  in addition to the  $A$, $\mu^\ast$,  $\gamma_1$,  $p_1^\ast$ and $p_2^\ast$ parameter, we have nuisance parameters which allow us to change
  the flux threshold of the point-source detection by a factor of $k=0.5$ to 2.0 relative to $S_{\rm thr}$ (more comments are provided at the end of this subsection).
  
\footnotetext{\label{foot::table_best_fit}There is a subtlety connected to the counting of 
                 the degrees of freedom (dof) in the fit of the blazar model to the 4FGL+4LAC catalog data.
                 We use the total number of point sources as upper limit and the number of BLL+BCU+FSRQ as lower 
                 limit in 7 flux bins.
                 The number of fit parameters is 5. Using this information gives a dof of 2. The subtlety is that, on top of 
                 the mentioned constraints, we use for some parameter points redshift information as lower limit in the fit, in effectively 28 bins. 
                 However, the $\chi^2$ at the best-fit point is only marginally
                 affected by these lower limits. So, we decided not to count this information in the dof stated in the table.}  
  
 The results are reported in Fig.~\ref{fig::fit_results_Cp}. The left-panel refers to the auto-correlation APS amplitude $C_P$ as a function of the energy, while the right panel stands for one case of cross-correlation, specifically the cross-correlation of the $[8.3, 14.5]$ GeV energy bin with all the other bins. We note that the best-fit model well reproduces the measurement obtained in Ref.  \cite{Ackermann:2018wlo}, demonstrating that the blazar model is compatible also with the APS of the photon field fluctuations, and that the study of the unresolved components by means of two different methods (the 1pPDF and the APS) provide consistent results, as quantified below. The best-fit values for the parameters and their errors are reported in Tab. \ref{tab::bestfit_parameter2}: the results are well compatible with those obtained in the 1pPDF analysis, including the value obtained for the $p_2^\ast$ parameter. While the 1pPDF and APS results are well compatible with the catalog results, the fact that $p_2^\ast$ turns out somehow lower for both analyses (sensitive to the unresolved blazar component) might be indicative that the fainter blazar emission starts to point toward a slightly different regime.

Previous analyses of gamma-ray APS found evidence for two populations instead of a single population \cite{Fornasa:2016ohl,Ando:2017alx,Ackermann:2018wlo}. We also test here this hypothesis, following a strategy already used in Ref.~\cite{Ando:2017alx}. On top of the blazar physical model, we add an additional soft and faint component for which we assume  $dN / dS = A_\mathrm{PWL} (S/S_0)^{-\beta_\mathrm{PWL}}$ (where $S$ refers to the flux in the energy bin 1--100 GeV) and an energy spectrum given by $d N/d E \sim E^{-\Gamma_\mathrm{PWL}}$. We then perform a fit with the sum of the blazar physical model plus such additional generic power-law component. In total, this fit involves 8 free parameters: the 5 parameters already used in our reference analysis, plus $A_\mathrm{PWL}$, $\beta_\mathrm{PWL}$, $\Gamma_\mathrm{PWL}$. We find a slight improvement in the $\chi^2$, but not statistically significant, being smaller than at the 2$\sigma$ C.L.
This then justifies the adopted procedure to fit the $C_P$ with a single blazar population: namely, the underlying assumption of our analysis that blazars are the dominant contributor to the unresolved gamma-ray sky, in the regime just below the {\em Fermi}-LAT detection threshold. Notice that we are adopting  a different approach as compared to Ref.~\cite{Ackermann:2018wlo}, where a preference for 2 populations was instead present: we describe the gamma ray emission in terms of a physical blazar model and we allow for a distribution of their spectral indices $\Gamma$ with a dispersion of $\sigma = 0.28$ \cite{Ajello:2015mfa} (see Eq. (\ref{eq:GLF_0})), instead of adopting a given spectral index as done in Ref.~\cite{Ackermann:2018wlo}. In this case,  the single-blazar model is able to describe the measured APS. We leave for a future work the investigation of the  possible presence of subdominant additional unresolved populations.
We just mention here that we found some degeneracy between the addition of a new population and the size of the parameter $\sigma$ in Eq.~\eqref{eq:GLF_0}. The latter tends to increase in the absence of a second population (with an upper limit at around 0.3).

\subsection{Complementarity of 1pPDF, $\mathbf{C_P}$ and 4FGL catalog}\label{sec::complementarity}

\begin{figure}[t]
  \includegraphics[width=1.0\linewidth, trim={2.0cm 0 1.9cm 0}, clip]{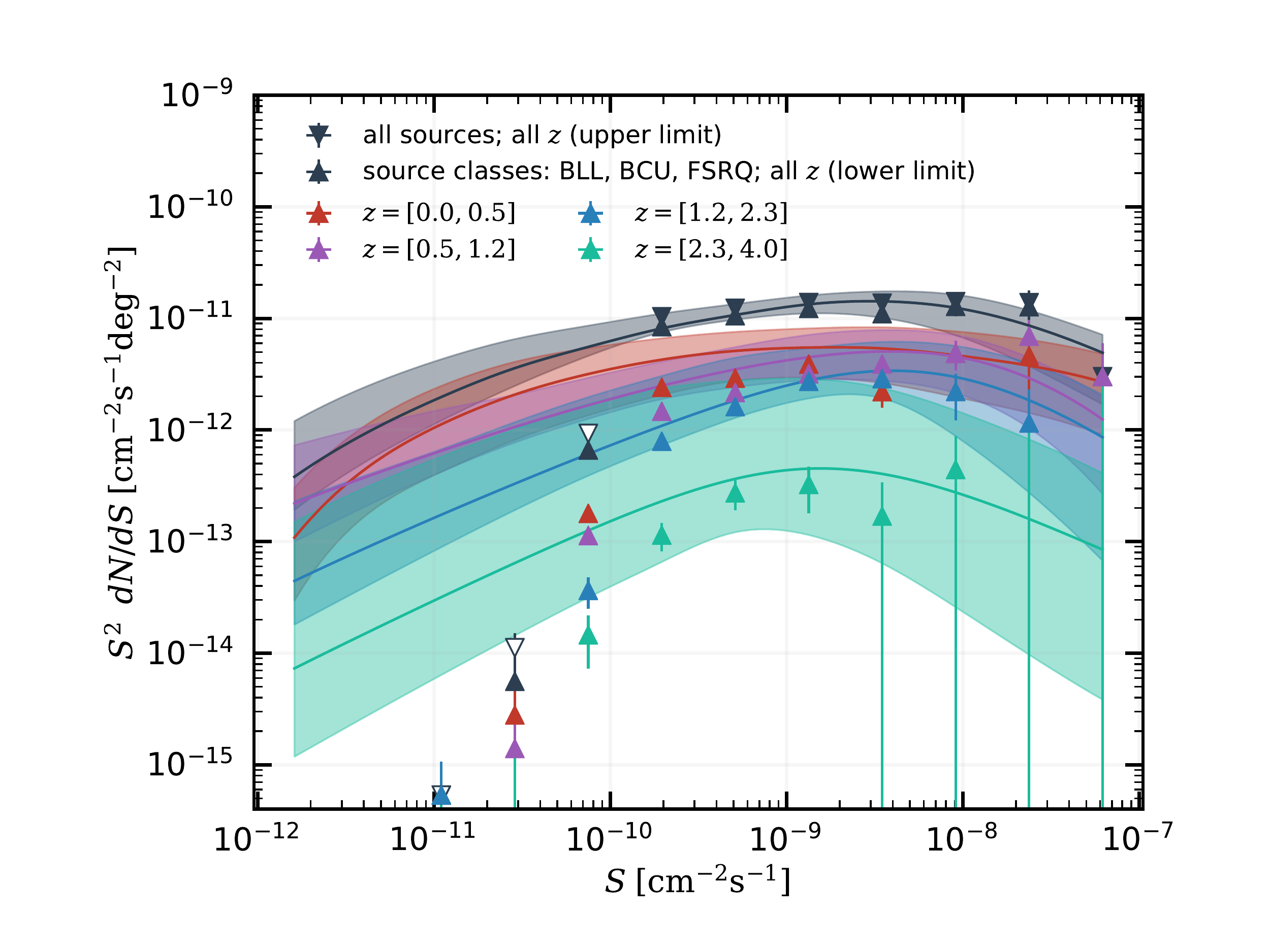}
  \caption{Comparison of the source count distributions extracted from the 4FGL and 4LAC catalogs (data points) with the
               best-fit blazar model of the 4FGL fit (solid lines). The shaded bands display the 1$\sigma$ uncertainty.
               Data points with triangles pointing upwards (downwards) 
               have to be understood as lower (upper) limits. The open white data points are below the flux threshold 
               and, therefore, not considered in the fit. The flux $S$ refers to the energy bin from 1~GeV to 100~GeV. 
  }
  \label{fig::fit_results_4FGL}
\end{figure}

\begin{center}
\begin{figure}[t]
       \includegraphics[width=1.0\linewidth, trim={2.0cm 0 1.9cm 0}, clip]{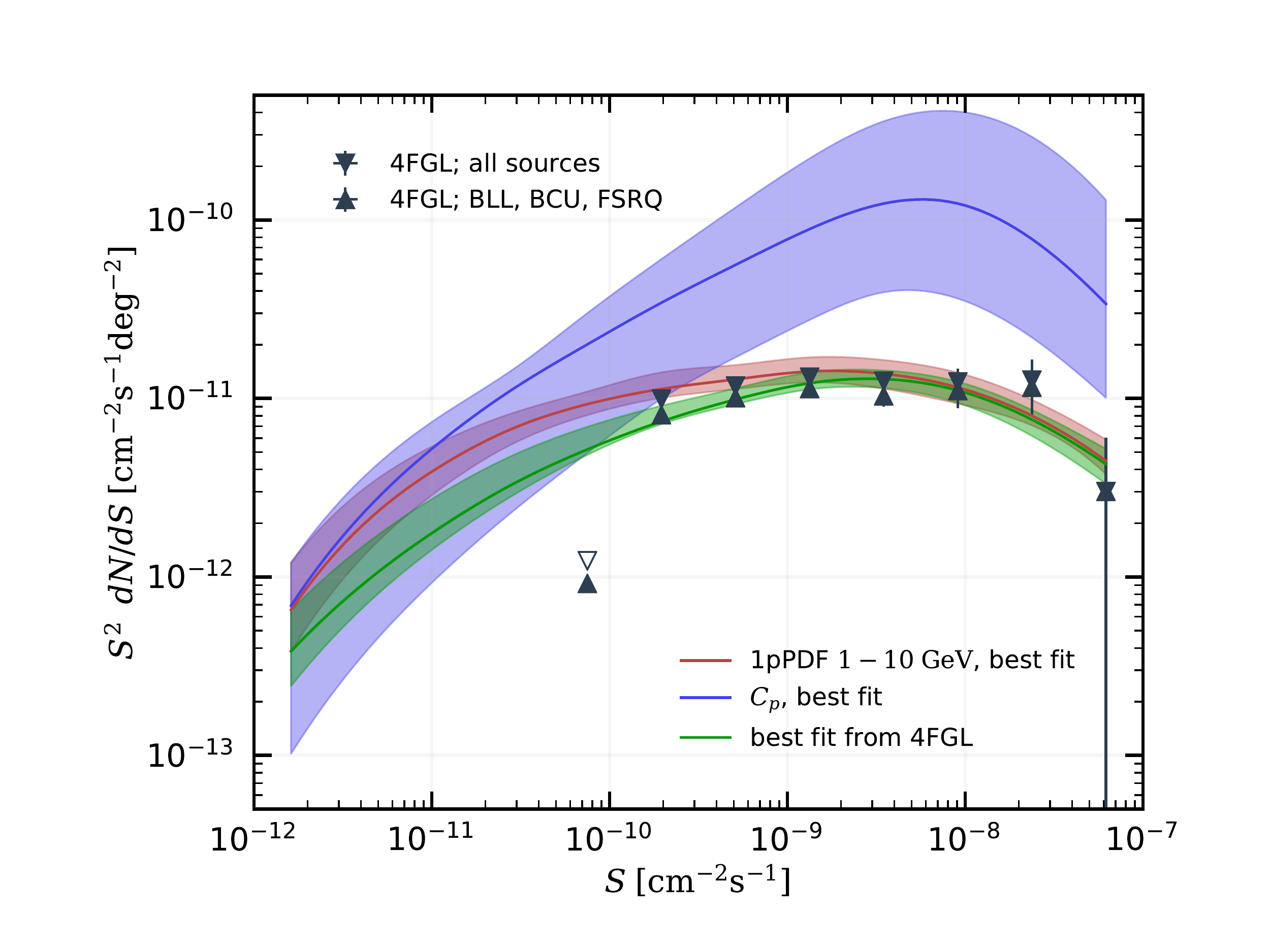}
    \caption{Source-count distribution $\diff N /\diff S$ in the energy bin from 1~GeV to 10~GeV, as obtained from the best- fits parameters arising from the fit of each each individual observable (1pPDF, APS and catalogs).
                  Solid lines refer to the best-fit values of the parameters, while the shaded bands give the corresponding 1$\sigma$
                  uncertainty. To guide the eye we add the $\diff N / \diff S$ points determined from the 4FGL catalog; 
                  lower triangles contain all source classes while upper triangles restrict to the 
                  source classes BBL, BCU, and FSRQ. 
                		         }
  \label{fig::predictions_dNdS_1_10}
\end{figure}
\end{center}

The two methods adopted to investigate the unresolved side of the gamma-ray emission (1pPDF and APS) produce compatible results, but also provide complementary information. This can be seen by analyzing the full parameter space, reproduced in Fig.~\ref{fig::triangle_1ppdf_Cp}, which shows the 1-dimensional and 2-dimensional $\chi^2$ distributions.

The preferred regions obtained with the two techniques always exhibit overlap within a 2$\sigma$ C.L, demonstrating compatibility. However,  the APS analysis significantly constrains the central value of the blazar spectral index $\mu^\ast$, while being much less effective on the other parameters. This occurs because the APS analysis involves several energy bins (through the cross-correlation in energy) and this allows us to characterize the blazar SED. On the other hand,  the 1pPDF method is more constraining on the other GLF parameters, especially the normalization $A$ and the parameter $\gamma_1$ which governs the luminosity evolution.  Clearly, since in the 1pPDF we are adopting a single energy bin, we have small sensitivity on the SED.

\begin{figure*}[t!]
	{\includegraphics[width=0.48\textwidth, trim={1.5cm 0 2.5cm 0}, clip]{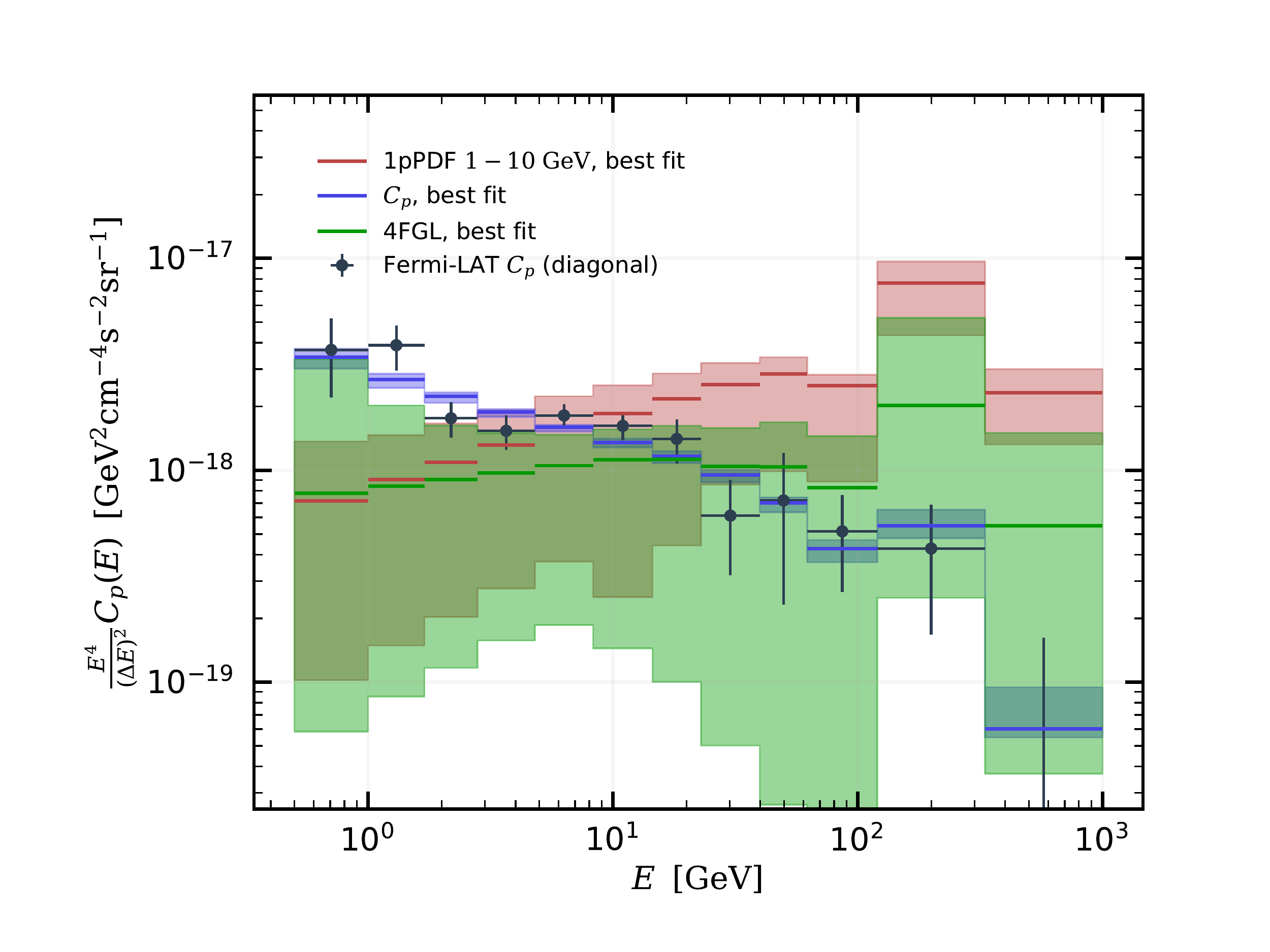}\hspace{0.04\textwidth}\includegraphics[width=0.48\textwidth, trim={1.5cm 0 2.5cm 0}, clip]{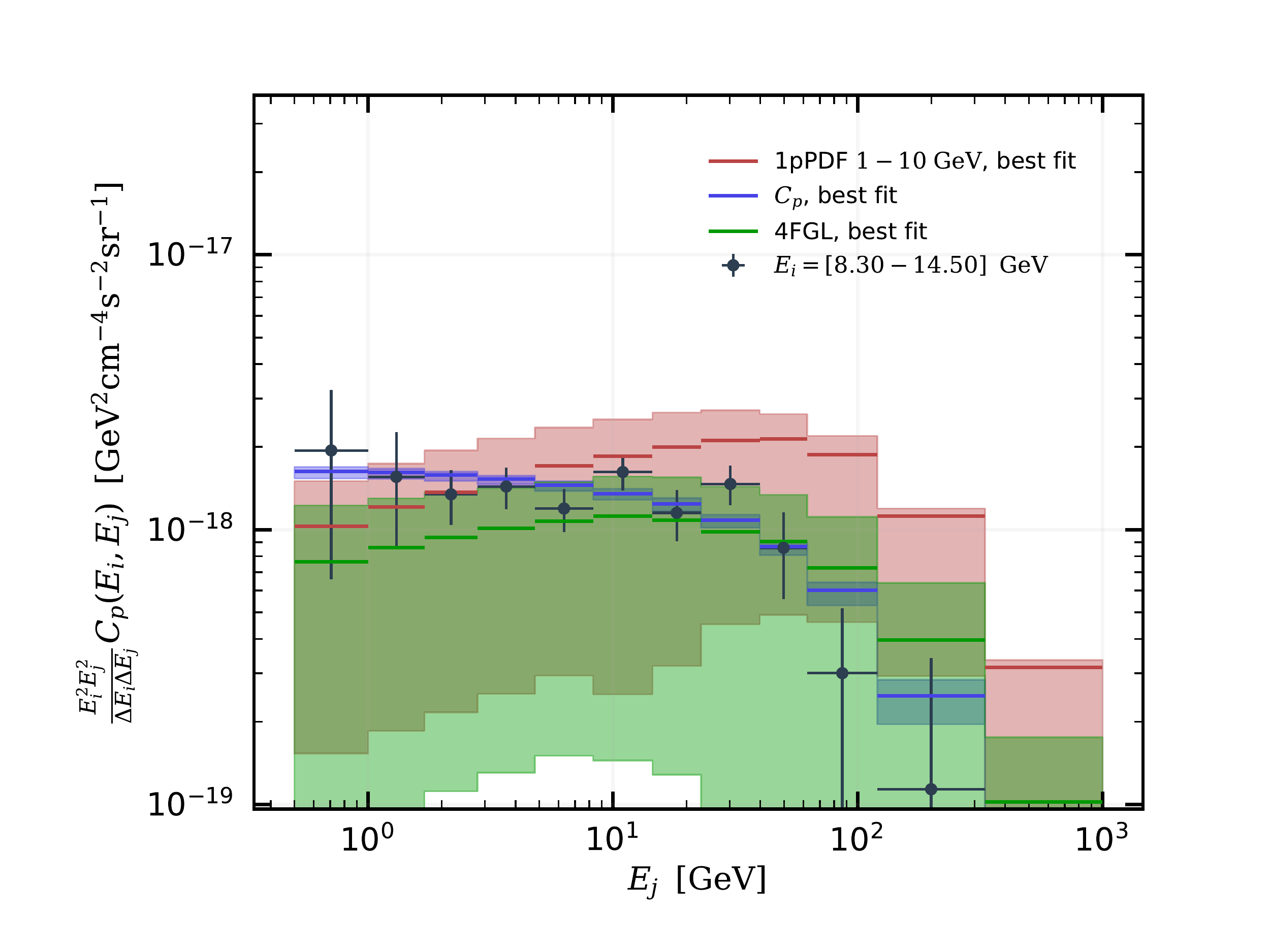}}
        \caption{ Amplitude of the angular correlation $C_P$ as arising from the various fits to the different observables (1pPDF, APS and catalogs).  Solid lines refer to the best-fit values of the parameters, while the shaded bands give the corresponding 1$\sigma$  uncertainty. The left panel refers to the autocorrelation (in energy), while the right panel show one set of cross-correlations (in energy) of one selected energy bin
	           (8.3~GeV--14.5~GeV) with all others. }
  \label{fig::predictions_Cp}
\end{figure*}
\begin{figure*}[t!]
	{\includegraphics[width=0.9\textwidth]{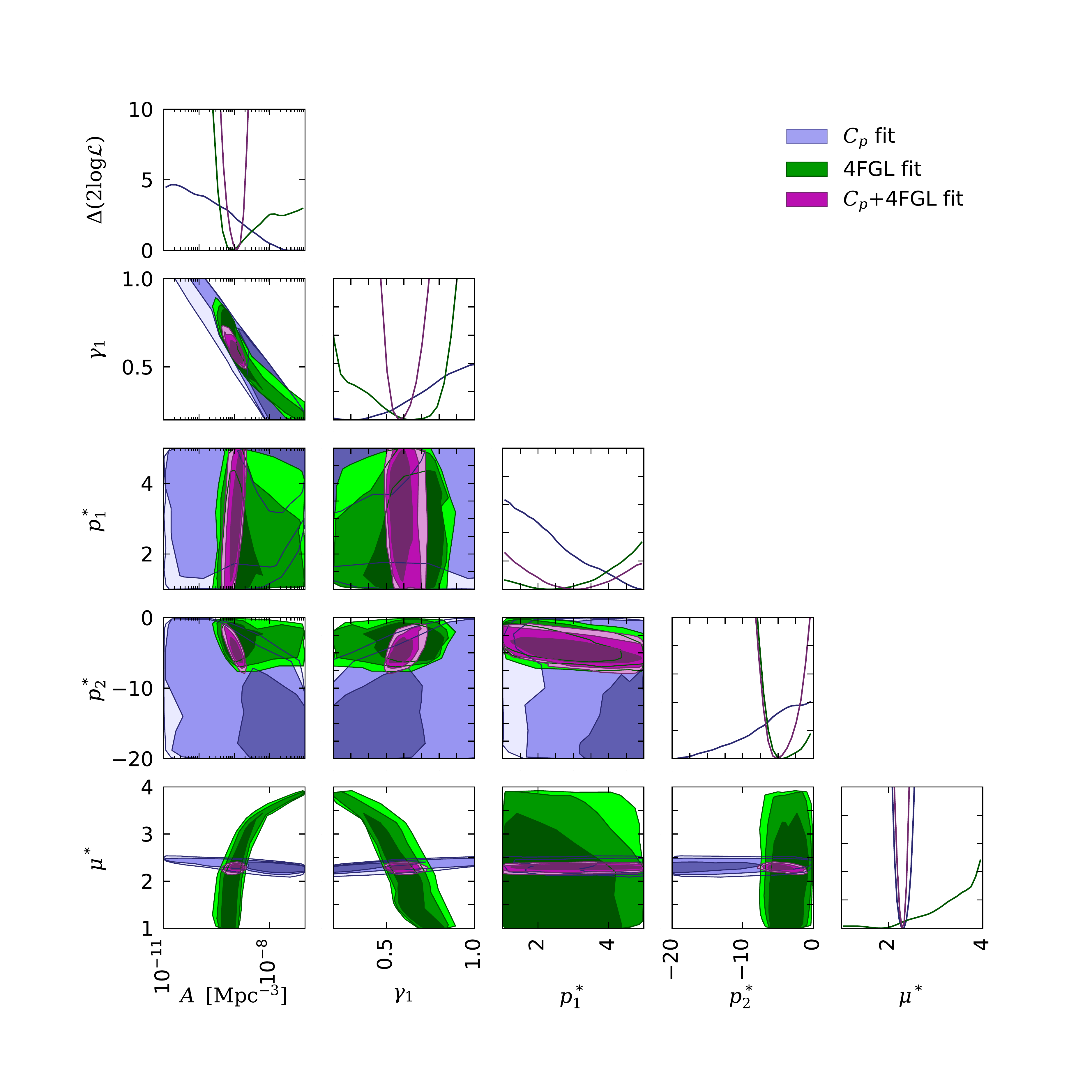}}
	\caption{ Comparison of the joint fit of $C_P$+4FGL with the individual fits of the $C_P$ and the 4FGL catalog, 
	          whose contours were already shown Fig.~\ref{fig::triangle_1ppdf_Cp}.  }
  \label{fig::triangle_Cp_4FGL}
\end{figure*}

The results of the blazar model fit to the 4FGL catalog are shown in Fig.~\ref{fig::fit_results_4FGL}. The lower and upper black triangles 
mark the source count distribution of all point sources $(dN/dS)_{\mathrm{all},i}$ and  point sources associated as blazars $(dN/dS)_{\mathrm{as},i}$, respectively.  The best fit of the blazar model lies between the two source count distributions, which serve as upper and lower limit in the fit. The colored triangles show the source count distribution in four redshift bins $(dN/dS)_{z,ij}$. Those data points are a lower limit to the blazar model, since the redshift catalog is incomplete. We observe that the best-fit model fulfills this requirement, and lies above the colored data points for all the redshift ranges.
The corresponding best fit parameters for this fit are reported in Tab.~\ref{tab::bestfit_parameter2}.

The results obtained by fitting the source count distribution of the 4FGL  catalog are also provided in Fig.~\ref{fig::triangle_1ppdf_Cp} (green contours). The results are well compatible with those obtained with the 1pPDF and APS analyses. As expected, there is  very good agreement to the 1pPDF analysis,  since the catalogs and the 1pPDF analysis directly probe the number of point sources, although in two different regimes (resolved for catalog, resolved and unresolved for 1pPDF). We note that the catalog fit provides the strongest constraints on the parameter $p_2^*$, by excluding values smaller than about $-7$. To interpret this constraint, we remind that $p_2^*$ 
changes the shape of the LDDE at $z\gsim z_c^*=1.25$. The other two methods cannot exclude small values of $p_2^*$ since, in contrast to the catalog fit, they do not contain explicit redshift information.

As a further result, we show in Figs.~\ref{fig::predictions_dNdS_1_10} and~\ref{fig::predictions_Cp} how the different observables would be reconstructed if only the best-fit from one of the techniques is used. 
In Fig.~\ref{fig::predictions_dNdS_1_10} we show that the source count distribution provided by the best-fit parameters of the APS analysis is in good agreement with the 1pPDF and 4FGL catalog analyses for what concerns the unresolved regime. On the other hand, the APS study would over-predict the measured \dnds\ in the resolved part. The lack of precision of the $C_P$ analysis in this regime is somewhat expected, since it is based only on data below {\em Fermi}-LAT detection threshold. If one attempts to describe a complete model of both resolved and unresolved blazars, this has to be complemented by other techniques, as we show at the end of this section.

The prediction that would be obtained for the APS as a function of energy by using only the information coming from the 1pPDF or from the 4FGL catalog analyses is shown in Fig.~\ref{fig::predictions_Cp}. Since they are obtained in a single energy bin, they cannot be very predictive for what concerns the blazar SED. This becomes manifest if one compares the precision obtained from the APS analysis (blue regions) in the reconstruction of the energy spectrum to what is predicted by the 1pPDF (red) or the 4FGL catalog (green) analyses. 
Therefore, Figs.~\ref{fig::predictions_dNdS_1_10} and~\ref{fig::predictions_Cp} reinstate the complementarity of the different probes in cornering the blazar model. 
We note that the prediction of the \dnds\ from the $C_P$ and vice versa show deviations above the 1$\sigma$ level. A similar deviation is visible also in the parameter contours shown in Fig.~\ref{fig::triangle_1ppdf_Cp}. We checked explicitly that at the 3$\sigma$ level all the bands are compatible with the data points.
We also checked explicitly the compatibility between the $C_P$ and 1pPDF predictions and the $dN/dS$ of the catalog in all our 4 redshift bins.

The contours from 1pPDF cannot be simply combined with APS or 4FGL analyses without computing the appropriate co-variance. Indeed, the 1pPDF uses data both in the resolved and unresolved regimes. The combination can be instead performed between APS and 4FGL analyses, since they rely on separate data-sets.    To demonstrate again the complementarity between the $C_P$ measurement and the information in the 4FGL catalog, we perform 
   a further joint fit to both observables, in which the joint $\chi^2_{C_P+\mathrm{4FGL}}$ is defined as sum of the two 
   individual $\chi^2$s defined in Eqs.~\eqref{eqn::chi2_ASP} and \eqref{eqn::chi2_4FGL}, respectively.  
We obtain a good fit with a  minimal joint $\chi^2$/dof of 90.9/79 which can be separated into a contribution from 
the $C_P$ fit of 86.6  and the 4FGL fit and 4.4.
The combination of both observables guarantees that both, the measured \dnds\ (Fig.~\ref{fig::predictions_dNdS_1_10}) in the resolved part and the measured $C_P$ (Fig.~\ref{fig::predictions_Cp}) in the unresolved regime, are properly reproduced at 1$\sigma$.
Furthermore, we observe that the nuisance parameter, $k$, 
is very well constrained by the combination of the two methods, since the 4FGL information fixes the \dnds\ above $S_\mathrm{thr}$.
As a further test for our treatment of the detection efficiency, we computed the predicted resolved flux for the model resulting from $C_P + {\rm 4FGL}$ fit. 
For each energy band used in the APS analysis, the resulting fluxes (normalized by the factor $E_1\,E_2/(E_2-E_1)$ where $E_1$ and $E_2$ are lower and upper bound of the energy band) are: $[3.24,$ 2.79, 2.44, 2.13, 1.83, 1.56, 1.25, 0.087, 0.067, 0.049, 0.023, $0.015]\times 10^{-7}$ GeV cm$^{-2}$ s$^{-1}$ sr$^{-1}$. We verified that these flux values are always lower than the sum of the fluxes of detected point sources in 4FGL, confirming that our threshold approximation leads to consistent results.

Results are shown in Fig.~\ref{fig::triangle_Cp_4FGL} and the best-fit values are reported into Tab.~\ref{tab::bestfit_parameter2}.
One can explicitly note the striking complementarity already mentioned above, namely, the best-fit regions shrink to the overlap of the two individual fits. We recommend to use the values of the $C_P$+4FGL fit to obtain a good agreement or the blazar model in the resolved and unresolved regime.

\section{Conclusions} \label{sec::conclusion}
In this paper we adopted and compared different statistical methods to constrain the gamma-ray emission from blazars. Based on the most recent \emph{Fermi}-LAT data at high Galactic latitudes, we derived the description of the blazar luminosity function and spectral energy distribution, with best-fit parameters provided in Tab.~\ref{tab::bestfit_parameter2}.

The global contribution of unresolved gamma-ray point sources to the EGB can be probed through the statistical properties of the observed gamma-ray counts. We analyzed the 1pPDF and two-point APS, and compared the results to the characterization provided by the analysis of resolved sources in the 4FGL catalog.
We found that the 1pPDF and APS can indeed extend our knowledge of the blazar GLF and SED to the unresolved regime, and are  able to determine the \dnds\ of blazars down to fluxes almost two orders of magnitude smaller than the \emph{Fermi}-LAT detection threshold for resolved sources.

The different approaches provide predictions that are generically in good agreement with each other.
Moreover, they show a significant complementarity.
The APS analysis better characterizes the blazar SED, since it involves several energy bins (and their cross-correlation).
The 1pPDF is more constraining for what concerns the normalization and luminosity evolution of the GLF. The analysis of the redshift distribution of the resolved sources in the catalogs allows a more refined determination of the GLF redshift evolution.
The complementarity of the different techniques in constraining the parameters of the GLF and SED models of blazars can be appreciated in Figs.~\ref{fig::triangle_1ppdf_Cp} and~\ref{fig::triangle_Cp_4FGL}.

Finally, we notice that, for the blazar gamma-ray luminosity function, there is an overall consistency between our best-fit parameters (reported in Tab.~\ref{tab::bestfit_parameter2}) and those obtained in Ref.~\cite{Ajello:2015mfa}, based on a previous version of the catalog of resolved sources. Especially when comparing our results obtained with the 4FGL catalog with those of Ref.~\cite{Ajello:2015mfa}, the values of the parameters are all well compatible. This seems to suggest that the additional sources identified in 4FGL basically share the same features of those brighter sources present the catalog adopted in Ref.~\cite{Ajello:2015mfa}. When information from the unresolved sources is added (anisotropies and 1pPDF analyses), some deviations arise,  especially for the redshift evolution parameters $p_1^*$  and $p_2^*$ (although with sizeable errors for the 1pPDF). This might be suggestive of a difference in redshift behaviour when approaching fainter sources, which are populating the unresolved sky. However, uncertainties are still large to make firm conclusions. When combining the $C_P$ and the 4FGL analyses, the parameters are consistent with those of Ref.~\cite{Ajello:2015mfa}, but better determined (smaller errors), the only exception being $\mu^*$, for which a $3 \sigma$ difference in its central values is found. This again might be indicative of a possible transition to a different regime.

We plan for future works to further constrain the GLF of blazars, and potentially other source populations (\textit{e.g.} mAGNs or SFG), by investigating the 1pPDF in different energy bins, and by performing a two-point correlation analysis with catalogs of blazars at different wavelengths.

\medskip
\section*{\label{sec::acknowledgments}Acknowledgments}
%
%
We wish to thank M. Di Mauro and M. Negro for fruitful discussions and advice.
This work  is supported by:  `Departments of Excellence 2018-2022' grant awarded by the Italian Ministry of Education, University and Research (\textsc{miur}) L.\ 232/2016; Research grant `The Anisotropic Dark
Universe' No.\ CSTO161409, funded by Compagnia di Sanpaolo and University of Turin; Research grant TAsP (Theoretical Astroparticle Physics) funded \textsc{infn}; Research grant `The Dark Universe: A Synergic Multimessenger Approach' No.\ 2017X7X85K funded by \textsc{miur};  Research grant ``Deciphering the high-energy sky via cross correlation'' funded by the agreement ASI-INAF n. 2017-14-H.0; Research grant ``From  Darklight  to  Dark  Matter: understanding the galaxy/matter connection to measure the Universe'' No. 20179P3PKJ funded by MIUR.
 
\bibliography{bibliography}{}
\bibliographystyle{apsrev4-1.bst}

\end{document}